\documentclass[sigconf]{acmart}

\usepackage{subcaption}
\usepackage{array, makecell, multirow, colortbl}
\usepackage{tikz}
\usetikzlibrary{trees}
\usepackage{cleveref}
\usepackage{nicefrac, enumitem, appendix, pifont}
\usepackage{listings}
\usepackage{capt-of} % for \captionof in non-floating environments
\usepackage{changepage}
\usepackage[ruled,vlined]{algorithm2e}
\usepackage{xspace}
\usepackage{lipsum}
\usepackage{cuted}
\usepackage[most]{tcolorbox}
\tcbuselibrary{skins,breakable}

\lstdefinestyle{evocode}{
  basicstyle=\ttfamily\footnotesize,
  columns=fullflexible,
  breaklines=true,
  breakatwhitespace=true,
  showstringspaces=false,
  tabsize=4,
  frame=single,
  framerule=0.4pt,
  rulecolor=\color{black!15},
  backgroundcolor=\color{gray!5},
  numbers=left,
  numberstyle=\tiny\color{black!45},
  xleftmargin=1.6em,
  framexleftmargin=1.6em,
  aboveskip=0.6em,
  belowskip=0.4em
}

\newtcolorbox{CaseStudyBox}{
    enhanced,
    breakable,
    colback=white,
    colframe=blue!80,      % 现在的边框颜色
    arc=1.5mm,             % 圆角
    boxrule=1pt,
    left=3mm, 
    right=3mm, 
    top=3mm,               % 建议稍微加大 top，因为没有标题栏后需要一点呼吸空间
    bottom=3mm,
}

\definecolor{impgreen}{RGB}{0,120,90}
\definecolor{impred}{RGB}{160,40,40}

\setcopyright{none}
\authorsaddresses{}
\renewcommand\footnotetextcopyrightpermission[1]{}
\acmConference[KDD'26]{KDD 2026}{August 9 to 13, 2026}{Jeju, Korea}

\newcommand{\projectname}{\textbf{\textsc{EvoCodeBench}}\xspace}
\begin{document}

%%
%% The "title" command has an optional parameter,
%% allowing the author to define a "short title" to be used in page headers.
\title{EvoCodeBench: A Human-Performance Benchmark for Self-Evolving LLM-Driven Coding Systems}

% Wentao Zhang, Jianfeng Wang, Liheng Liang, Yilei Zhao, HaiBin Wen, Zhe Zhao

\author{
    Wentao Zhang$^{1,\dagger}$ $\mathpunct{,}$ Jianfeng Wang$^{2,\dagger}$ $\mathpunct{,}$ Liheng Liang$^{3,\dagger}$ $\mathpunct{,}$ Yilei Zhao$^{1}$ $\mathpunct{,}$ HaiBin Wen$^{4}$ $\mathpunct{,}$ Zhe Zhao$^{5,\ddagger}$ \\ 
    {
        \fontsize{10}{10}\selectfont
        \textsuperscript{\rm 1} Nanyang Technological University 
        \textsuperscript{\rm 2} East China University of Science and Technology \\ 
        \textsuperscript{\rm 3} Guangdong Ocean University 
        \textsuperscript{\rm 4} City University of Hong Kong 
        \textsuperscript{\rm 5} Stanford University \\ 
        zhangwent963@gmail.com, ehz0002@stanford.edu 
    } 
}

\renewcommand{\shortauthors}{Wentao Zhang et al.}

\begin{abstract}

% 会自我提升，自动进化的动态benchmark

% 首先，目前很多的工作都开始focus到agent的自进化，无论是算法进化、prompt进化等，都是在考虑提升agent的解决任务的能力以及效率，但是没有bench来衡量agent的自进化的能力的提升。尤其是在推理过程中进化，解决任务准确率、效率等是否在逐渐提升的，目前的固定静态的benchmark完全无法体现这一点。

% 次之，目前市场的bench都没有考虑与人类水平对比（打败了多少人），或者说衡量agent/llm的智力是人类的多少倍，打败多少人这样的指标可以对比

% 最后，需要满足多语言，目前很多benchmark并没有详细对比多种编程语言差异，尤其是小语种，例如市场上python占比超过了22%，而小语种例如kotlin只有0.97%，能否在这种差异下LLM具备一定的coding稳定性是值得考察的方向。

% 关于这3点motivation如果abstract，introduction，related work缺少了也要适当地补上（这三点都是重要的，尤其前两点，必须要要重视）

As large language models (LLMs) continue to advance in programming tasks, LLM-driven coding systems have evolved from one-shot code generation into complex systems capable of iterative improvement during inference. However, existing code benchmarks primarily emphasize static correctness and implicitly assume fixed model capability during inference. As a result, they do not capture inference-time self-evolution, such as whether accuracy and efficiency improve as an agent iteratively refines its solutions. They also provide limited accounting of resource costs and rarely calibrate model performance against that of human programmers. Moreover, many benchmarks are dominated by high-resource languages, leaving cross-language robustness and long-tail language stability underexplored. Therefore, we present \projectname, a benchmark for evaluating self-evolving LLM-driven coding systems across programming languages with direct comparison to human performance. \projectname tracks performance dynamics, measuring solution correctness alongside efficiency metrics such as solving time, memory consumption, and improvement algorithmic design over repeated problem-solving attempts. To ground evaluation in a human-centered reference frame, we directly compare model performance with that of human programmers on the same tasks, enabling relative performance assessment within the human ability distribution. Furthermore, \projectname supports multiple programming languages, enabling systematic cross-language and long-tail stability analyses under a unified protocol. Our results demonstrate that self-evolving systems exhibit measurable gains in efficiency over time, and that human-relative and multi-language analyses provide insights unavailable through accuracy alone. \projectname establishes a foundation for evaluating coding intelligence in evolving LLM-driven systems.

\end{abstract}

%%
%% The code below is generated by the tool at http://dl.acm.org/ccs.cfm.
%% Please copy and paste the code instead of the example below.
%%
% \begin{CCSXML}
% <ccs2012>
%  <concept>
%   <concept_id>00000000.0000000.0000000</concept_id>
%   <concept_desc>Do Not Use This Code, Generate the Correct Terms for Your Paper</concept_desc>
%   <concept_significance>500</concept_significance>
%  </concept>
%  <concept>
%   <concept_id>00000000.00000000.00000000</concept_id>
%   <concept_desc>Do Not Use This Code, Generate the Correct Terms for Your Paper</concept_desc>
%   <concept_significance>300</concept_significance>
%  </concept>
%  <concept>
%   <concept_id>00000000.00000000.00000000</concept_id>
%   <concept_desc>Do Not Use This Code, Generate the Correct Terms for Your Paper</concept_desc>
%   <concept_significance>100</concept_significance>
%  </concept>
%  <concept>
%   <concept_id>00000000.00000000.00000000</concept_id>
%   <concept_desc>Do Not Use This Code, Generate the Correct Terms for Your Paper</concept_desc>
%   <concept_significance>100</concept_significance>
%  </concept>
% </ccs2012>
% \end{CCSXML}

% \ccsdesc[500]{Do Not Use This Code~Generate the Correct Terms for Your Paper}
% \ccsdesc[300]{Do Not Use This Code~Generate the Correct Terms for Your Paper}
% \ccsdesc{Do Not Use This Code~Generate the Correct Terms for Your Paper}
% \ccsdesc[100]{Do Not Use This Code~Generate the Correct Terms for Your Paper}

%%
%% Keywords. The author(s) should pick words that accurately describe
%% the work being presented. Separate the keywords with commas.
%% A "teaser" image appears between the author and affiliation
%% information and the body of the document, and typically spans the
%% page.

%%
%% This command processes the author and affiliation and title
%% information and builds the first part of the formatted document.
%% Kill the deferred footnote BEFORE \maketitle queues author addresses
%%
\maketitle

\begingroup
\renewcommand{\thefootnote}{\fnsymbol{footnote}}
\footnotetext[2]{These authors contributed equally.}
\footnotetext[3]{Corresponding author. Email: zhangwent963@gmail.com, ehz0002@stanford.edu}
\endgroup

\section{Introduction}
Large language models (LLMs) have progressed from code-completion utilities to general-purpose programming assistants and building blocks for agentic coding systems. In contrast to one-shot code generation, contemporary agents can \emph{self-evolve} during inference by iteratively refining candidate programs, repairing failures using execution feedback, and reusing intermediate artifacts, tools, or memory to improve both correctness and efficiency within the same problem-solving process. This shift motivates a central evaluation challenge: how should coding competence be measured when behavior is adaptive at inference time and performance is jointly determined by solution quality, resource cost, and improvement dynamics? Early benchmarks such as HumanEval~\citep{chen2021evaluating}, MBPP~\citep{austin2021program}, and APPS~\citep{hendrycks2021apps} standardized evaluation by measuring functional correctness via unit tests under fixed specifications, while competition-level datasets such as CodeContests~\citep{li2022alphacode} further raised the difficulty ceiling. Subsequent work strengthened test suites~\citep{liu2024evalplus}, increased freshness via continuously updated algorithmic tasks (LiveCodeBench~\citep{jain2024livecodebench}), and addressed repository-level software engineering settings (SWE-Bench~\citep{jimenez2023swe}), while recent extensions broaden coverage across languages and incorporate efficiency-oriented signals such as execution cost and resource overhead~\citep{raihan2025mhumaneval,xu2025cruxeval,qing2025effibench,huang2024effibench}. However, a coherent framework for evaluating inference-time self-evolution in coding agents remains underdeveloped.

Despite these advances, existing benchmarks exhibit three limitations when applied to LLM-based coding systems, including one-shot models and agentic systems that self-evolve during inference. First, the prevailing paradigm remains static and benchmarks treat problems as independent one-shot instances and score systems primarily by final correctness, which obscures inference-time dynamics. Recent methods such as Self-Refine~\citep{madaan2023selfrefine} and Reflexion~\citep{shinn2023reflexion} demonstrate that iterative refinement can substantially boost pass rates, yet existing benchmarks rarely quantify whether a system becomes more accurate and efficient within the inference process, how quickly it improves as it refines solutions, or what time and memory costs are incurred, so they cannot distinguish fast, reliable solvers from systems that succeed only after expensive trial-and-error~\citep{olausson2024selfrepair}. Second, results are typically reported in absolute accuracy units with limited calibration to human performance. Without human-referenced metrics, it is difficult to interpret what a given pass rate implies about practical coding competence or to report where a system falls on a human-calibrated scale, including its percentile rank or the number of humans it outperforms on the same tasks. Third, multilingual coverage is often dominated by a single language and ecosystem (often Python), which can mask cross-language generalization and bias evaluation toward familiar tooling and idioms. Although multilingual benchmarks such as MultiPL-E~\citep{cassano2023multiple} have extended evaluation to many languages, they still lack efficiency and human-referential metrics. This imbalance is especially problematic under long-tail usage patterns. According to the TIOBE Index for January 2026, Python accounts for 22.61\% and Kotlin accounts for 0.97\%~\citep{tiobe2026}. Such market skew can translate into differences in the amount of training data available per language, potentially inducing systematic bias in LLMs, yet existing benchmarks rarely measure or diagnose this effect.

We introduce \projectname to address these gaps with an evaluation framework for LLM-driven coding systems, including both one-shot models and self-evolving agents. \projectname evaluates systems on competitive-programming-style problems in multiple mainstream languages (e.g., Python, C++, Java, Go, and Kotlin), and it is designed to capture inference-time improvement rather than only final outcomes. Beyond pass rate, \projectname records efficiency and resource usage signals, including wall-clock solving time, memory consumption, and improvement trajectories as the system solves problems during inference. We pair these measurements with platform-side judging in a standardized execution environment, which returns acceptance status and fine-grained performance feedback such as runtime, memory usage, and percentile-based \emph{beats} statistics, enabling analyses of whether gains in correctness come with disproportionate cost. We also provide standardized evaluation scripts and reporting templates to facilitate reproducibility and fair comparisons across model families and agent configurations. Finally, \projectname enables human-referenced reporting by situating model performance within the distribution of human submissions on the same tasks, including percentile ranks and the number of human solutions a system outperforms. Specifically, we make the following contributions:
\begin{itemize}[leftmargin=*,topsep=0pt,itemsep=0pt,parsep=0pt]
    \item We present the first systematic, human-referenced evaluation of LLM-based coding systems, situating model performance within the empirical distribution of human submissions and enabling interpretable comparisons in terms of percentile rank and the number of humans a system outperforms.
    \item We reformulate coding evaluation from a static, endpoint-focused paradigm to a trajectory-level analysis that jointly characterizes correctness, efficiency, and resource cost during inference-time self-evolution, and we empirically validate the necessity and stability of this evaluation framework across multiple languages, including long-tail programming ecosystems.
    \item We release \projectname, a large-scale multilingual benchmark with standardized judging, human baselines, and reproducible evaluation protocols, and demonstrate how its metrics provide actionable insights for the design and future development of coding LLMs and agentic systems.
\end{itemize}

\vspace{-0.3cm}
\section{Related Work}
\subsection{Evaluating LLMs on Isolated Coding Tasks}

Early benchmarks for evaluating LLMs on programming tasks primarily emphasize static correctness on small, self-contained problems. Representative datasets such as HumanEval~\citep{chen2021evaluating} and MBPP~\citep{austin2021program} assess whether a generated function satisfies a predefined specification by passing a fixed suite of unit tests, while APPS~\citep{hendrycks2021apps} and CodeContests~\citep{li2022alphacode} raise the difficulty ceiling to competitive-programming-level challenges with thousands of problems. Subsequent variants with strengthened test suites, including HumanEval Pro and MBPP Pro~\citep{yu2024humaneval} and EvalPlus~\citep{liu2024evalplus}, further tighten correctness criteria to better discriminate among strong models. These benchmarks have played a central role in standardizing evaluation protocols for functional code generation and tracking rapid progress in model capabilities.

However, these evaluations largely assume fixed model behavior during inference and treat each problem as an independent one-shot instance. Consequently, they provide limited visibility into inference-time self-evolution, even though methods such as Self-Refine~\citep{madaan2023selfrefine} and Reflexion~\citep{shinn2023reflexion} have shown that iterative refinement can substantially improve pass rates, and analyses of self-repair~\citep{olausson2024selfrepair} highlight the need to account for computational cost when evaluating such gains. To address concerns about data contamination and task realism, later benchmarks such as LiveCodeBench~\citep{jain2024livecodebench} curate continuously updated algorithmic problems sourced from competitive programming platforms and evaluate auxiliary skills such as self-repair and test-output prediction. Despite improvements in data freshness and diversity, the dominant paradigm remains centered on final correctness on fixed tasks and offers limited calibration to human performance or systematic characterization of inference-time efficiency dynamics.

\subsection{Benchmarking LLM-Driven Coding Systems}

Beyond isolated function synthesis, recent work has increasingly shifted toward evaluating LLM-based agents and more complex coding systems in realistic software engineering settings. Repository-level benchmarks assess a model’s ability to resolve real-world issues through multi-file code edits, dependency reasoning, and test-driven validation, marking an important step toward capturing practical workflows~\citep{jimenez2023swe,liu2023repobench}. Class-level benchmarks such as ClassEval~\citep{du2024classeval} move beyond function-level evaluation to assess generation of interdependent code units, while broader task suites like BigCodeBench~\citep{zhuo2024bigcodebench} evaluate diverse function-call composition across libraries. In parallel, agent-oriented benchmarks investigate multi-step reasoning, tool use, and interaction~\citep{liu2023agentbench,guo2024stabletoolbench,mialon2023gaia,yao2024tau}, and multi-agent frameworks such as MapCoder~\citep{islam2024mapcoder} demonstrate that decomposing the code generation pipeline into specialized agents can improve competitive-programming performance.

More recent extensions expand evaluation along additional dimensions such as multilingual coverage and efficiency awareness by examining cross-language generalization~\citep{cassano2023multiple,ding2023crosscodeeval,raihan2025mhumaneval,xu2025cruxeval} and incorporating execution cost, runtime performance, or memory overhead into evaluation protocols~\citep{huang2024effibench,qing2025effibench}. However, these extensions primarily add new axes without fundamentally revising the underlying evaluation target. Even when efficiency signals are reported, they are rarely integrated with inference-time adaptation and improvement dynamics, and multilingual coverage is often dominated by high-resource languages, leaving long-tail stability underexplored.

\subsection{Limitations of Existing Benchmarks}
Across both isolated coding benchmarks and agent-oriented evaluation frameworks, a common limitation persists. Even as state-of-the-art models approach performance saturation on strengthened algorithmic benchmarks (e.g., HumanEval Pro and MBPP Pro) and demonstrate competitive results on realistic software engineering tasks (e.g., SWE-Bench subsets), there is no established framework that jointly measures inference-time self-evolution, efficiency, and human-relative capability. In particular, existing benchmarks provide limited human-referenced metrics, which constrains the interpretability of claims about human-level or superhuman coding ability. Moreover, they do not explicitly characterize whether systems become faster or more accurate during inference as they refine solutions~\citep{madaan2023selfrefine,shinn2023reflexion}, nor do they systematically link such improvements to time and memory costs. Finally, language coverage remains skewed toward high-resource languages, and market-level usage skew can translate into training-data imbalance and language bias \citep{tiobe2026}, yet benchmarks rarely quantify this effect. These gaps motivate the need for an evaluation framework that jointly models performance \textbf{dynamics}, \textbf{efficiency}, \textbf{human-relative capability}, and \textbf{cross-language stability}, which we address in \projectname.

% 首先，目前很多的工作都开始focus到agent的自进化，无论是算法进化、prompt进化等，都是在考虑提升agent的解决任务的能力以及效率，但是没有bench来衡量agent的自进化的能力的提升。尤其是在推理过程中进化，解决任务准确率、效率等是否在逐渐提升的，目前的固定静态的benchmark完全无法体现这一点。

% 次之，目前市场的bench都没有考虑与人类水平对比（打败了多少人），或者说衡量agent/llm的智力是人类的多少倍，打败多少人这样的指标可以对比

% 最后，需要满足多语言，目前很多benchmark并没有详细对比多种编程语言差异，尤其是小语种，例如市场上python占比超过了22%，而小语种例如kotlin只有0.97%，能否在这种差异下LLM具备一定的coding稳定性是值得考察的方向。

% 关于这3点motivation如果abstract，introduction，related work缺少了也要适当地补上（这三点都是重要的，尤其前两点，必须要要重视）

\begin{figure*}[htbp]
    \centering
    \begin{minipage}[t]{0.69\textwidth}
        \centering
        \includegraphics[width=\linewidth]{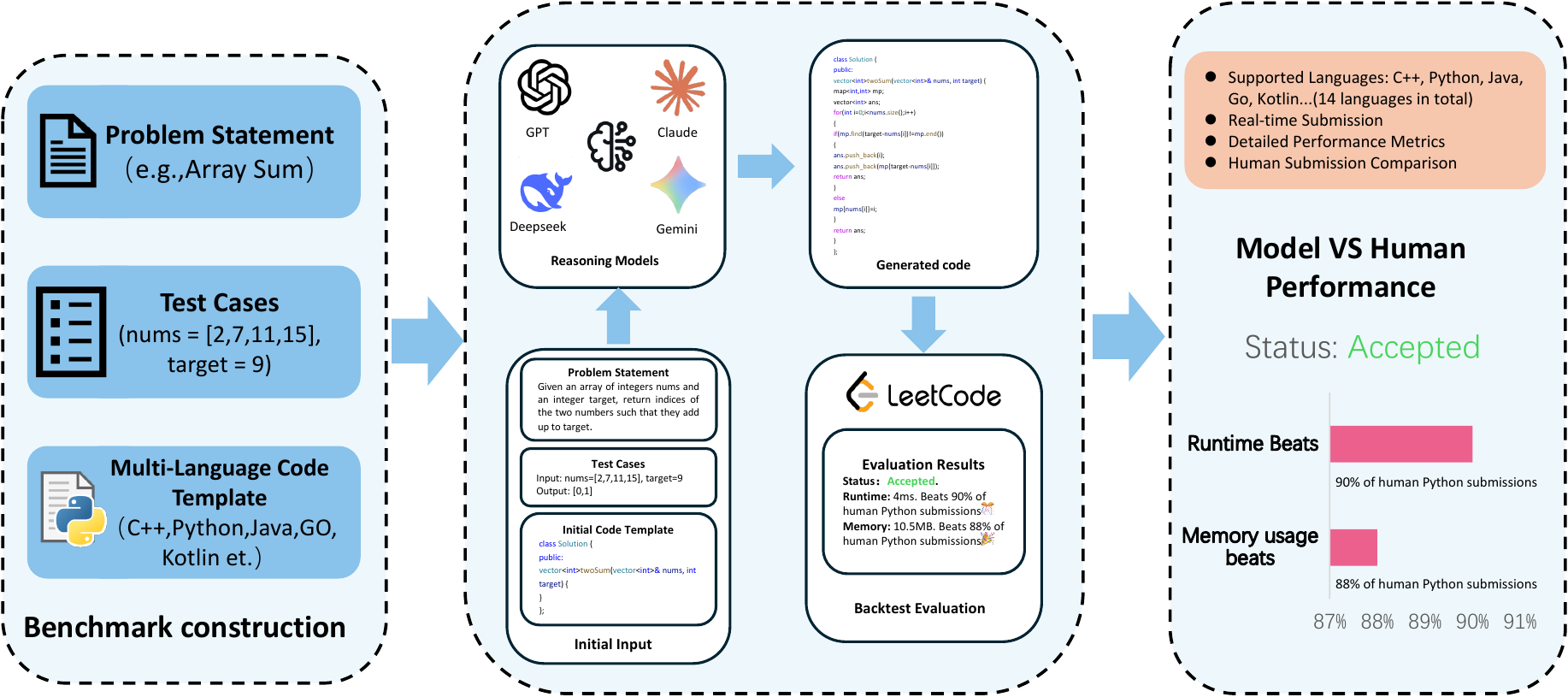}
        \caption{Code evaluation pipeline.}
        \label{fig:code-evaluation-pipeline}
    \end{minipage}\hfill
    \begin{minipage}[t]{0.29\textwidth}
        \centering
        \includegraphics[width=\linewidth]{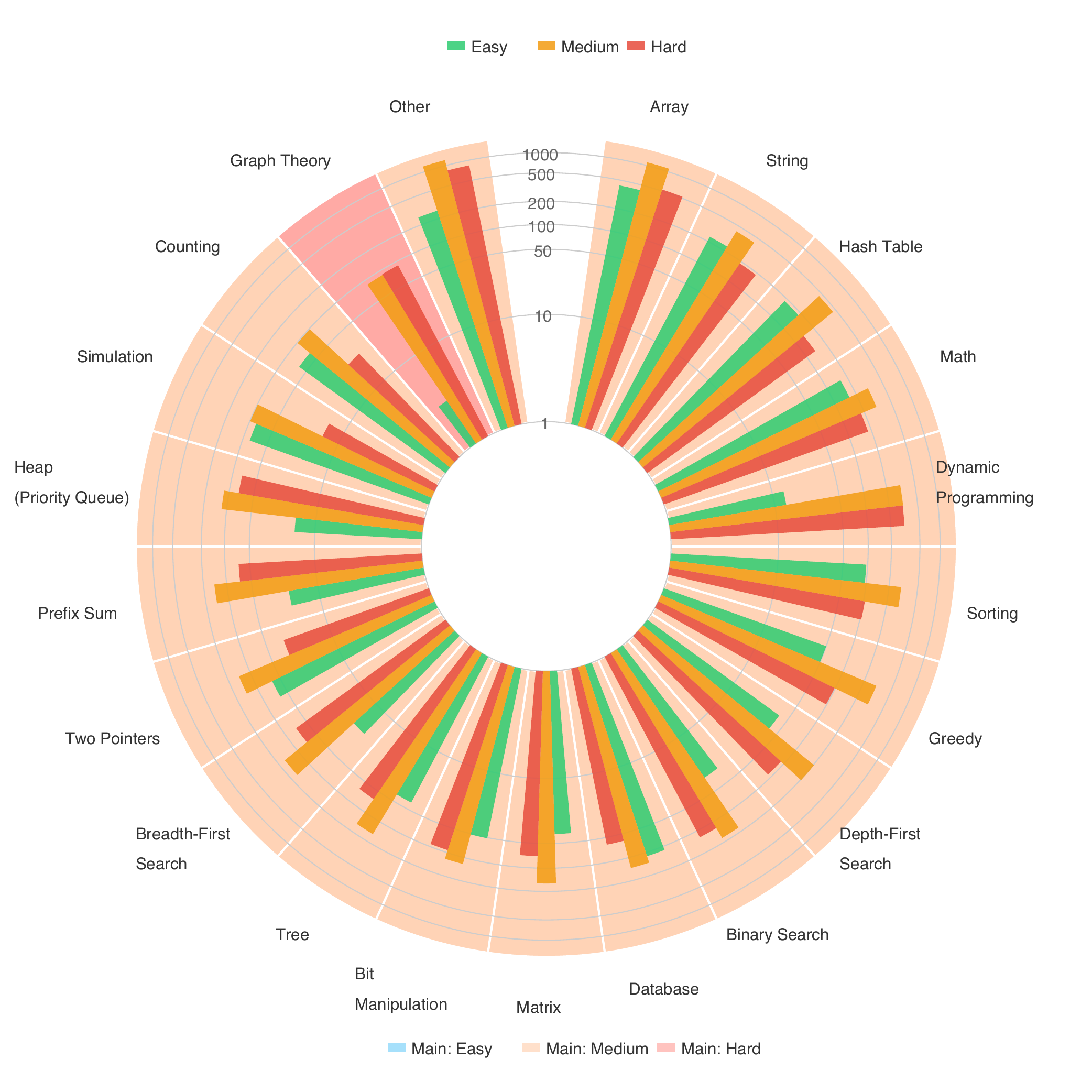}
        \caption{Distribution of problems.}
        \label{fig:leetcode-tags-distribution}
    \end{minipage}
    \vspace{-0.6cm}
\end{figure*}

\section{EvoCodeBench}

\subsection{Benchmark Design Rationale}

Our benchmark construction is driven by the three motivations outlined above, namely evaluating inference-time self-evolution, calibrating performance against human ability, and assessing cross-language robustness under long-tail language usage. To this end, we build \projectname on top of the LeetCode online judge, which provides an execution-based evaluation interface and rich feedback signals that closely match our requirements. Specifically, acceptance status and pass rate, together with the number of passed test cases for partially correct submissions, enable fine-grained assessment of functional correctness beyond binary success. For accepted submissions, the platform reports runtime and memory usage, along with percentile-based runtime beats and memory beats statistics computed against the distribution of human submissions, which directly supports human-referenced evaluation. Finally, LeetCode provides standardized starter code across many programming languages, enabling consistent multi-language evaluation and the study of long-tail language stability under a unified measurement protocol. These signals also facilitate analyzing efficiency--accuracy trade-offs and identifying failure modes that emerge only under constrained runtime or memory budgets.

\subsection{Benchmark Construction}

\textbf{Data Collection.} We collect problems through the SQL interface provided by LeetCode and retrieve the full set of 3,822 programming problems available at the time of crawling. For each problem, we extract the full natural-language statement and normalize it into a structured representation that preserves key semantic elements, including embedded images when present, as well as the official input--output examples. We store image links and resolve them to local assets so that models can be evaluated with the same visual context as human solvers. In addition, we collect language-specific starter code templates for a broad set of languages, enabling consistent prompt construction and fair cross-language comparison. Each problem is further annotated with its platform-provided difficulty label (Easy, Medium, and Hard) and a set of topical tags that describe the required algorithmic concepts and data structures, such as arrays, trees, dynamic programming, mathematics, and strings. During ingestion, we perform basic quality checks, including filtering malformed records, removing duplicates, and validating that problem statements, examples, and templates are successfully parsed. Figure~\ref{fig:leetcode-tags-distribution} summarizes tag and difficulty distributions, used to guide dataset balancing and analyze performance variation across problem categories and difficulty strata.

\textbf{Problem Characteristics.}
LeetCode-style algorithmic problems provide a controlled yet demanding setting for evaluating coding competence in both one-shot LLMs and LLM-based agents. Each task specifies explicit constraints and input--output behavior that must be matched exactly, stressing instruction following, edge-case coverage, and implementation fidelity under a fixed interface. The breadth of tags and difficulty levels further tests algorithm selection, data-structure proficiency, and complexity-aware reasoning, while requiring robust translation from high-level ideas to executable code. Because evaluation is execution-based, brittle solutions are reliably exposed through concrete failures such as off-by-one errors, corner-case bugs, and language-specific pitfalls. Standardized starter templates across languages enable systematic cross-language comparisons, including robustness in long-tail languages with distinct syntax and runtime characteristics. Finally, since many problems admit within-inference refinement, this setting probes agentic behaviors such as self-repair, hypothesis testing against examples, and efficiency-aware optimization under runtime and memory budgets.

\textbf{Problem Evaluation.}
We adopt a unified evaluation protocol that standardizes model interaction with the execution environment through a lightweight \textbf{vanllia coding agent}, enabling fair and reproducible comparison across foundation models. For each problem instance, the agent presents a fixed input representation to the model and deterministically extracts executable source code from the model’s output. The code is then submitted to an official execution-based judging service, which evaluates functional correctness on hidden test cases and reports resource usage statistics. This design ensures that all models are assessed under identical input formats, execution conditions, and scoring criteria.

To capture inference-time self-evolution, we further introduce a \textbf{self-evolving coding agent} that allows models to iteratively refine their solutions within a single problem-solving episode. Under a fixed revision budget, the agent supports multiple rounds of solution updates while keeping the task specification, prompt structure, and evaluation interface unchanged. This controlled setup enables direct comparison between one-shot generation and inference-time adaptive refinement, isolating the effect of self-evolution from confounding factors such as prompt variation or task reformulation.

\subsection{Agent Construction}

In the preceding section, we introduced a lightweight \textbf{coding agent} and a \textbf{self-evolving coding agent} as two standardized evaluation interfaces for comparing foundation models on \projectname. In this subsection, we detail how each agent is constructed, including the prompting and output-format constraints, deterministic code extraction, and the submission-and-feedback loop with the online judge. We then describe the within-inference reflection--revision procedure used by the self-evolving agent, enabling analysis under the same task specification and scoring protocol.

\textbf{Coding Agent.} 
As illustrated in Appendix~\ref{appx_sec:case-add-two-numbers}, the coding agent is deliberately designed to be lightweight and minimally augmented. It receives as input a standardized prompt consisting of the problem statement and a language-specific starter template, and it instructs the model to produce a format-constrained response with two explicit fields. The \texttt{reasoning} field contains the model’s intermediate reasoning, while the \texttt{code} field contains the final solution expressed in the provided template format. The agent deterministically extracts the \texttt{code} field and submits it to the online judge for execution-based evaluation. Importantly, this baseline agent operates in a single-pass manner and does not incorporate external memory, tool use, or iterative refinement. Both the prompting schema and the output parsing rules are fixed across models, ensuring that observed performance differences can be attributed to the underlying model capabilities rather than agent-level augmentation or inference-time adaptation.

\textbf{Self-evolving Coding Agent.} 
To evaluate inference-time self-evolution, we design a self-evolving coding agent that performs controlled, within-inference refinement of solution code. Starting from an initial model-generated draft, the agent iteratively improves the solution through a fixed-budget sequence of reflection and revision steps within a single problem-solving episode. At each iteration, the agent prompts the model to analyze its current solution and identify potential failure modes, including unhandled corner cases, logical errors, suboptimal complexity, or language-specific issues. Based on this analysis, the model proposes targeted code revisions aimed at improving correctness or efficiency. The updated solution is then resubmitted to the online judge, and execution feedback is recorded. Crucially, only the solution artifact is modified during self-evolution. The problem specification, prompt schema, evaluation interface, and model parameters remain fixed, ensuring that observed gains are attributable solely to inference-time refinement rather than changes in task formulation or external augmentation. By constraining refinement to a single inference episode with a fixed revision budget, this agent enables systematic analysis of improvement trajectories in both correctness and efficiency.

\section{Empirical Studies}

\subsection{Experimental Settings}

\textbf{Benchmark Settings.} 
We construct the evaluation set by sampling 100 problems from the full collection of 3,822 tasks, spanning diverse tag categories to cover a broad range of algorithmic skills. We avoid evaluating on the full pool because widely circulated LeetCode problems, particularly older ones, may overlap with the training data of commonly used LLMs. Such overlap can inflate measured performance and confound comparisons through memorization. We therefore focus on a recent set of 100 problems as a held out test set to reduce the risk of data leakage while preserving topical diversity. We evaluate models in five programming languages. We include widely used languages such as Python3, C++, Java, and Go, and we additionally include Kotlin as a representative long tail language. This choice enables analysis of cross-language robustness under skewed language popularity and quantifies performance shifts when moving from high- to lower-resource settings.

\textbf{LLMs Settings.} 
We evaluate a set of frontier models, including \texttt{deepseek-v3.2}, \texttt{grok-4.1-fast}, \texttt{gemini-3-flash-preview}, \texttt{gemini-3-pro-preview}, \texttt{claude-sonnet-4.5}, \texttt{claude-opus-4.5} and \texttt{gpt-5.2}. For all models, we enable reasoning by default. The maximum number of generated tokens to 65,536 for each response.

\textbf{Agent Settings.} 
We use the coding agent without additional agent level constraints beyond the standardized prompting and parsing protocol. For the self-evolving coding agent, we allow at most three reflection and revision iterations per problem.

\subsection{Evaluation Metrics}

We report three groups of metrics that capture complementary aspects of coding performance. First, capability metrics measure functional correctness and failure modes under the judge constraints. Second, efficiency metrics summarize runtime and memory cost for accepted submissions. Third, human metrics quantify how often accepted submissions outperform the distribution of human solutions in runtime and memory.

\noindent\textbf{Coding capability metrics.}
\begin{itemize}[leftmargin=*,topsep=0pt,itemsep=0pt,parsep=0pt]
    \item \textbf{Pass Rate (PR)}: The number of problems that the model successfully passed all test cases within the time and memory limits.
    \item \textbf{Time Limit Exceeded (TLE)}: The number of problems where the model's solution exceeded the allowed execution time limit.
    \item \textbf{Memory Limit Exceeded (MLE)}: The number of problems where the model's solution exceeded the allowed memory usage.
    \item \textbf{Compile Error (CE)}: The number of problems where the model's generated code failed to compile due to syntax errors.
    \item \textbf{Runtime Error (RE)}: The number of problems where the model's solution encountered a runtime error during execution.
    \item \textbf{Wrong Answer (WA)}: The number of problems where the model's solution produced incorrect output for the test cases.
    \item \textbf{Timeout (TO)}: The number of problems where the model failed to generate a response within the specified timeout period.
    \item \textbf{Response Error (RpE)}: The number of problems where the model failed to return a valid or parseable code response.
\end{itemize}
\noindent\textbf{Coding efficiency metrics.}
\begin{itemize}[leftmargin=*,topsep=0pt,itemsep=0pt,parsep=0pt]
    \item \textbf{Average Runtime (AR)}: The average execution runtime in milliseconds of the model's accepted solutions across all problems.
    \item \textbf{Average Memory (AM)}: The average memory consumption in megabytes of the model's accepted solutions across all problems.
    \item \textbf{Average Passed Cases (APC)}: The average number of test cases that the model's solutions successfully passed before failure.
\end{itemize}

\noindent\textbf{Human-referenced metrics.}
\begin{itemize}[leftmargin=*,topsep=0pt,itemsep=0pt,parsep=0pt]
    \item \textbf{Average Runtime Beats (ARB)}: The percentage of accepted solutions where the model's runtime performance beats other human submissions.
    \item \textbf{Average Memory Beats (AMB)}: The percentage of accepted solutions where the model's memory beats human submissions.
\end{itemize}

\begin{table*}[t]
\centering
\footnotesize
\caption{Model performance comparison on \projectname.}
\vspace{-0.45cm}
\label{tab:model-performance}
\setlength{\tabcolsep}{6pt}
\renewcommand{\arraystretch}{0.2}
\begin{tabular}{l cccccccc ccc c >{\columncolor{cyan!15}}c}
\toprule
\textbf{Model} & \multicolumn{8}{c}{\textbf{Capability metrics}} & \multicolumn{3}{c}{\textbf{Efficiency metrics}} & \multicolumn{2}{c}{\textbf{Human metrics}} \\
\cmidrule(lr){2-9}\cmidrule(lr){10-12}\cmidrule(lr){13-14}
 & \textbf{PR} & \textbf{TLE} & \textbf{MLE} & \textbf{CE} & \textbf{RE} & \textbf{WA} & \textbf{TO} & \textbf{RpE} & \makecell{\textbf{AR (ms)}} & \makecell{\textbf{AM (MB)}} & \makecell{\textbf{APC}} & \makecell{\textbf{ARB (\%)}} & \multicolumn{1}{c}{\textbf{AMB (\%)}} \\

\midrule
\rowcolor{gray!15} \multicolumn{14}{c}{\textit{\textbf{Python3}}} \\
deepseek-v3.2 & \cellcolor{cyan!15}34 & 1 & 1 & 0 & 23 & 8 & 1 & 36 & \cellcolor{cyan!15}1806.79 & 55.91 & 640.38 & \cellcolor{cyan!15}63.04 & 25.81 \\
grok-4.1-fast & \cellcolor{cyan!15}73 & 9 & 0 & 0 & 3 & 13 & 0 & 3 & \cellcolor{cyan!15}1860.90 & 56.02 & 741.60 & \cellcolor{cyan!15}49.92 & 30.15 \\
claude-4.5-sonnet & \cellcolor{cyan!15}42 & 37 & 2 & 0 & 0 & 10 & 8 & 1 & \cellcolor{cyan!15}880.98 & 45.16 & 702.64 & \cellcolor{cyan!15}61.06 & 22.12 \\
claude-4.5-opus & \cellcolor{cyan!15}82 & 9 & 0 & 0 & 0 & 5 & 3 & 1 & \cellcolor{cyan!15}1559.87 & 70.77 & 749.45 & \cellcolor{cyan!15}64.77 & 32.70 \\
gemini3-flash-preview & \cellcolor{cyan!15}79 & 4 & 0 & 0 & 2 & 14 & 1 & 0 & \cellcolor{cyan!15}1376.19 & 56.59 & 750.89 & \cellcolor{cyan!15}73.28 & 36.62 \\
gemini-3-pro-preview & \cellcolor{cyan!15}79 & 3 & 2 & 0 & 10 & 6 & 0 & 0 & \cellcolor{cyan!15}1112.49 & 57.59 & 750.72 & \cellcolor{cyan!15}68.70 & 34.75 \\
gpt-5.2 & \cellcolor{cyan!15}84 & 7 & 0 & 0 & 4 & 4 & 1 & 0 & \cellcolor{cyan!15}1644.21 & 47.91 & 746.26 & \cellcolor{cyan!15}68.06 & 37.98 \\

\midrule
\rowcolor{gray!15} \multicolumn{14}{c}{\textit{\textbf{C++}}} \\
deepseek-v3.2 & \cellcolor{cyan!15}11 & 1 & 0 & 30 & 0 & 6 & 4 & 43 & \cellcolor{cyan!15}158.73 & 163.59 & 605.82 & \cellcolor{cyan!15}73.11 & 74.05 \\
grok-4.1-fast & \cellcolor{cyan!15}79 & 9 & 0 & 1 & 0 & 5 & 2 & 4 & \cellcolor{cyan!15}428.32 & 223.68 & 748.61 & \cellcolor{cyan!15}58.57 & 46.67 \\
claude-4.5-sonnet & \cellcolor{cyan!15}41 & 42 & 2 & 0 & 1 & 9 & 2 & 3 & \cellcolor{cyan!15}379.68 & 179.86 & 710.59 & \cellcolor{cyan!15}56.17 & 50.84 \\
claude-4.5-opus & \cellcolor{cyan!15}85 & 6 & 0 & 0 & 0 & 6 & 1 & 2 & \cellcolor{cyan!15}382.45 & 184.22 & 758.21 & \cellcolor{cyan!15}64.06 & 55.58 \\
gemini-3-flash-preview & \cellcolor{cyan!15}84 & 2 & 0 & 2 & 1 & 10 & 0 & 1 & \cellcolor{cyan!15}266.04 & 168.93 & 743.31 & \cellcolor{cyan!15}68.02 & 59.24 \\
gemini-3-pro-preview & \cellcolor{cyan!15}78 & 3 & 0 & 4 & 2 & 12 & 1 & 0 & \cellcolor{cyan!15}2500 & 173.56 & 760.28 & \cellcolor{cyan!15}70.08 & 63.50 \\
gpt-5.2 & \cellcolor{cyan!15}90 & 2 & 1 & 0 & 2 & 4 & 1 & 0 & \cellcolor{cyan!15}228.62 & 177.76 & 766.97 & \cellcolor{cyan!15}71.87 & 56.48 \\

\midrule
\rowcolor{gray!15} \multicolumn{14}{c}{\textit{\textbf{Java}}} \\
deepseek-v3.2 & \cellcolor{cyan!15}11 & 0 & 0 & 47 & 1 & 1 & 5 & 35 & \cellcolor{cyan!15}72.91 & 143.63 & 481.45 & \cellcolor{cyan!15}57.37 & 32.71 \\
grok-4.1-fast & \cellcolor{cyan!15}73 & 5 & 0 & 5 & 0 & 12 & 1 & 4 & \cellcolor{cyan!15}227.45 & 136.80 & 746.23 & \cellcolor{cyan!15}52.98 & 41.97 \\
claude-4.5-sonnet & \cellcolor{cyan!15}41 & 40 & 1 & 1 & 1 & 15 & 0 & 1 & \cellcolor{cyan!15}161.49 & 130.54 & 679.41 & \cellcolor{cyan!15}58.04 & 46.22 \\
claude-4.5-opus & \cellcolor{cyan!15}87 & 4 & 1 & 0 & 0 & 6 & 1 & 1 & \cellcolor{cyan!15}188.63 & 134.27 & 748.63 & \cellcolor{cyan!15}59.54 & 55.61 \\
gemini3-flash-preview & \cellcolor{cyan!15}84 & 0 & 0 & 2 & 2 & 9 & 1 & 2 & \cellcolor{cyan!15}125.04 & 126.09 & 752.86 & \cellcolor{cyan!15}71.03 & 59.18 \\
gemini-3-pro-preview & \cellcolor{cyan!15}81 & 2 & 0 & 3 & 0 & 10 & 4 & 0 & \cellcolor{cyan!15}132.53 & 140.21 & 745.75 & \cellcolor{cyan!15}71.00 & 52.75 \\
gpt-5.2 & \cellcolor{cyan!15}91 & 2 & 0 & 1 & 0 & 5 & 1 & 0 & \cellcolor{cyan!15}160.32 & 125.48 & 754.31 & \cellcolor{cyan!15}66.09 & 60.56 \\

\midrule
\rowcolor{gray!15} \multicolumn{14}{c}{\textit{\textbf{Go}}} \\
deepseek-v3.2 & \cellcolor{cyan!15}7 & 0 & 0 & 39 & 0 & 0 & 1 & 53 & \cellcolor{cyan!15}112.71 & 12.59 & 709.57 & \cellcolor{cyan!15}62.73 & 54.36 \\
grok-4.1-fast & \cellcolor{cyan!15}69 & 3 & 0 & 16 & 0 & 4 & 3 & 5 & \cellcolor{cyan!15}194.90 & 23.26 & 755.43 & \cellcolor{cyan!15}66.83 & 62.44 \\
claude-4.5-sonnet & \cellcolor{cyan!15}44 & 41 & 0 & 0 & 0 & 13 & 0 & 2 & \cellcolor{cyan!15}222.64 & 19.71 & 712.55 & \cellcolor{cyan!15}57.09 & 53.32 \\
claude-4.5-opus & \cellcolor{cyan!15}84 & 5 & 0 & 0 & 0 & 9 & 0 & 2 & \cellcolor{cyan!15}162.50 & 19.95 & 744.45 & \cellcolor{cyan!15}72.91 & 63.00 \\
gemini3-flash-preview & \cellcolor{cyan!15}82 & 1 & 0 & 9 & 0 & 7 & 0 & 1 & \cellcolor{cyan!15}139.22 & 22.01 & 739.46 & \cellcolor{cyan!15}76.22 & 63.48 \\
gemini-3-pro-preview & \cellcolor{cyan!15}73 & 2 & 0 & 8 & 0 & 14 & 1 & 2 & \cellcolor{cyan!15}138.64 & 27.73 & 748.78 & \cellcolor{cyan!15}74.98 & 66.38 \\
gpt-5.2 & \cellcolor{cyan!15}90 & 2 & 0 & 4 & 0 & 4 & 0 & 0 & \cellcolor{cyan!15}162.16 & 23.26 & 748.60 & \cellcolor{cyan!15}75.14 & 64.05 \\

\midrule
\rowcolor{gray!15} \multicolumn{14}{c}{\textit{\textbf{Kotlin}}} \\
deepseek-v3.2 & \cellcolor{cyan!15}7 & 2 & 0 & 0 & 0 & 0 & 1 & 48 & \cellcolor{cyan!15}59.29 & 62.27 & 793.57 & \cellcolor{cyan!15}58.33 & 74.56 \\
grok-4.1-fast & \cellcolor{cyan!15}62 & 2 & 0 & 22 & 0 & 8 & 2 & 4 & \cellcolor{cyan!15}307.45 & 75.45 & 759.55 & \cellcolor{cyan!15}78.12 & 72.83 \\
claude-4.5-sonnet & \cellcolor{cyan!15}42 & 36 & 1 & 8 & 1 & 10 & 1 & 1 & \cellcolor{cyan!15}192.62 & 78.49 & 757.64 & \cellcolor{cyan!15}81.59 & 77.79 \\
claude-4.5-opus & \cellcolor{cyan!15}83 & 4 & 0 & 5 & 0 & 5 & 0 & 3 & \cellcolor{cyan!15}210.47 & 76.60 & 750.98 & \cellcolor{cyan!15}83.18 & 76.53 \\
gemini3-flash-preview & \cellcolor{cyan!15}75 & 2 & 0 & 8 & 1 & 10 & 2 & 2 & \cellcolor{cyan!15}171.99 & 72.80 & 760.43 & \cellcolor{cyan!15}83.49 & 79.07 \\
gemini-3-pro-preview & \cellcolor{cyan!15}82 & 2 & 0 & 6 & 0 & 8 & 0 & 2 & \cellcolor{cyan!15}163.35 & 79.42 & 758.10 & \cellcolor{cyan!15}87.48 & 75.87 \\
gpt-5.2 & \cellcolor{cyan!15}78 & 1 & 0 & 17 & 0 & 4 & 0 & 0 & \cellcolor{cyan!15}149.06 & 75.13 & 766.42 & \cellcolor{cyan!15}88.44 & 83.12 \\

\bottomrule
\end{tabular}
\vspace{-0.3cm}
\end{table*}

\begin{figure*}[htbp]
    \centering
    \includegraphics[width=0.9\linewidth]{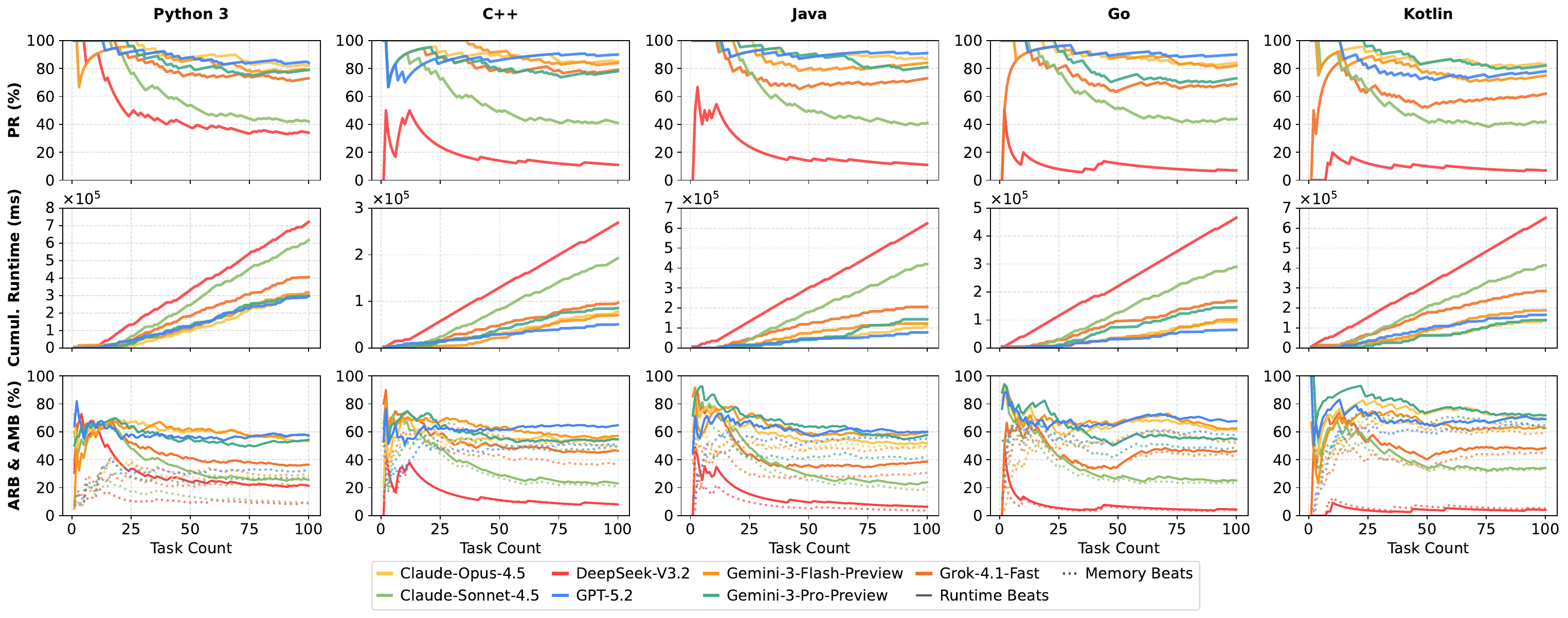}
    \vspace{-0.5cm}
    \caption{Performance comparison of models within-inference. Cumulative runtime is the total accumulated solving time where lower values indicate higher efficiency and time limit violations incur penalties.}
    \label{fig:inference-performance-comparison}
    \vspace{-0.5cm}
\end{figure*}

\subsection{Overall Comparison of Models}
Table~\ref{tab:model-performance} and Figure~\ref{fig:inference-performance-comparison} report the main results of \projectname across languages. The results are analyzed along three complementary dimensions: coding capability, efficiency, and human-referenced comparison. The empirical analysis are as follows:

\subsubsection{Comparison of Coding Capability}

We begin by analyzing pass rates together with detailed error profiles across programming languages, which jointly provide a diagnostic view of both solution quality and language-level execution robustness under heterogeneous constraints. In Figure~\ref{fig:inference-performance-comparison}, this dimension corresponds to the first row, which plots within-inference pass-rate trajectories as problems accumulate.

% 中文翻译：模型从高资源语言到长尾语言的表现会系统性下降，说明对语言覆盖的鲁棒性仍然有限。
\textbf{Finding 1. Model performance degrades systematically from high-resource to long-tail languages, indicating limited robustness to language coverage.}
\texttt{gpt-5.2} consistently emerges as the strongest performer in high-resource settings, achieving the highest pass rates in Python3, C++, Java, and Go, with PR values of 84, 90, 91, and 90, respectively. Notably, it is the only model to exceed a pass rate of 90 in Java. Figure~\ref{fig:inference-performance-comparison} corroborates this advantage in the first row by showing that \texttt{gpt-5.2} sustains one of the highest PR trajectories across these languages throughout the task sequence. However, its performance degrades in the long-tail language Kotlin, where the pass rate drops to 78. This degradation coincides with a sharp rise in compilation errors, with CE increasing to 17. More broadly, the first-row trajectories show that nearly all models experience a downward shift when moving from high-resource languages to Kotlin, indicating a structural robustness gap that persists even for top performers.

% 中文翻译：第二梯队模型的退化主要来自代码层面的错误，而非算法理解不足。
\textbf{Finding 2. Performance degradation in second-tier models is primarily driven by code-level errors rather than algorithmic misunderstanding.}
The three models \texttt{claude-4.5-opus}, \texttt{gemini-3-flash-preview}, and \texttt{gemini-3-pro-preview} form a competitive cluster that achieves pass rates primarily in the 70 to high-80 range across languages. Their performance gap relative to \texttt{gpt-5.2} is small in Python3 but becomes more pronounced in compiled languages. In these settings, compilation errors increase noticeably, particularly in Go and Kotlin. Importantly, wrong-answer rates within this tier remain relatively stable across languages, which suggests that the dominant bottleneck is not a wholesale collapse in algorithm selection. Instead, failures are better explained by language-specific implementation friction, such as syntax details, library and API usage, and toolchain constraints that prevent otherwise plausible solutions from executing correctly. This highlights that in compiled and long-tail languages, executability and correctness are tightly coupled, and code-level robustness can dominate observed performance even when high-level solution ideas are similar.

% 中文翻译：仅在高资源语言上评测会显著高估部分模型的真实编码能力。
\textbf{Finding 3. Evaluations restricted to high-resource languages can substantially overestimate the true coding capability of some models.}
\texttt{grok-4.1-fast} maintains pass rates in the high 60 to low 70 for Python3, C++, and Java, but shows a marked decline in Go and Kotlin, with pass rates dropping to 69 and 62, respectively. This decline is accompanied by elevated compilation error rates of 16 in Go and 22 in Kotlin. These results support the hypothesis that long-tail languages expose weaknesses in code generation robustness that are not apparent when evaluation is restricted to Python-centric benchmarks.

% 中文翻译：算法效率不足是一类独立的失败模式，单靠通过率难以刻画。
\textbf{Finding 4. Algorithmic inefficiency emerges as a distinct failure mode that is not captured by pass rate alone.}
The model \texttt{claude-4.5-sonnet} demonstrates consistently low pass rates across all languages, hovering around the low 40. In contrast to other models, its failures are dominated by time limit exceeded errors rather than compilation or response errors, with TLE counts exceeding 35 in every language evaluated. This indicates that the model frequently produces syntactically valid and executable code, but struggles to select or implement algorithms that meet the required efficiency constraints. Such behavior underscores the importance of explicitly separating functional correctness from algorithmic efficiency in the evaluation of coding systems, since execution-based benchmarks must account not only for whether a solution is correct, but also for whether it meets the computational constraints for practical deployment.

% 中文翻译：严重的输出格式违规暴露了结构化代码生成能力的缺陷。
\textbf{Finding 5. Severe output format violations in some models reveal weaknesses in structured code generation.}
The model \texttt{deepseek-v3.2} performs worst overall, exhibiting severe response errors across all languages. RpE counts exceed 35 in every setting and are further compounded by substantial compilation errors in C++, Java, and Go. These failures undermine execution-based evaluation and reveal weaknesses in output format compliance and instruction adherence, independent of algorithmic difficulty.

In conclusion, our results highlight that multilingual robustness, efficiency under judge constraints, and structured output reliability are all first-order factors for LLM-based coding systems. To improve performance, models should strengthen long-tail language competence through targeted multilingual training and toolchain-aware code generation that reduces compilation and execution-blocking errors. They should incorporate explicit efficiency reasoning and optimization to avoid time-limit failures, and adopt schema-constrained decoding and validation to prevent response-format violations that preclude execution-based evaluation. These directions are complementary and directly motivated by the failure patterns exposed by \projectname, reflecting practical limitations observed across languages and execution settings.

\subsubsection{Comparison of Coding Efficiency}
We next examine efficiency using the three columns in Table~\ref{tab:model-performance} that summarize average runtime, average memory, and average passed cases. These metrics quantify the computational cost of accepted solutions and the depth of progress when solutions fail. In Figure~\ref{fig:inference-performance-comparison}, the second row plots cumulative runtime as tasks accumulate. Lower curves indicate stronger end-to-end efficiency under the judge interface because time limit exceeded outcomes incur penalties and therefore increase accumulated cost.

% 中文翻译：端到端效率差异主要由解的可执行性稳定程度驱动。
\textbf{Finding 1. Differences in end-to-end efficiency are primarily driven by the stability of solution executability.}
In the second row of Figure~\ref{fig:inference-performance-comparison}, \texttt{gpt-5.2} consistently maintains the lowest or near-lowest cumulative runtime across all languages. This reflects not only fast accepted solutions but, more importantly, a high success rate with very few penalties from compilation or execution failures. Models such as \texttt{gemini-3-pro-preview}, \texttt{gemini-3-flash-preview}, and \texttt{claude-4.5-opus} initially track closely to the best-performing curve, but as tasks accumulate, particularly in Go and Kotlin, cumulative differences gradually emerge due to slightly lower executability, which results in more frequent failure penalties. In contrast, \texttt{claude-4.5-sonnet} and \texttt{deepseek-v3.2} show much steeper cumulative runtime growth, dominated by timeout penalties, even when some solutions are relatively fast. These observations indicate that stable end-to-end efficiency is primarily determined by the model’s ability to reliably produce executable solutions, and cumulative runtime effectively amplifies differences in executability across models.

% 中文翻译：即便在高通过率模型中，已通过样本上的条件运行时间与内存仍存在显著实现层差异。
\textbf{Finding 2. Among high pass-rate models, conditional runtime and memory still exhibit substantial implementation-level differences.}
Average runtime and average memory are computed only over accepted submissions and therefore characterize conditional efficiency among solved problems. Among higher pass-rate models, \texttt{gpt-5.2} achieves competitive runtimes in compiled languages, with 228.62 ms in C++, 160.32 ms in Java, and 162.16 ms in Go, while \texttt{gemini-3-pro-preview} attains the lowest average runtime among the stronger models in Python3 at 1112.49 ms. Memory exhibits a different structure. Java and C++ typically have larger memory footprints while Go is often lower and Kotlin falls in between, reflecting runtime-system differences. Within a fixed language, however, models still vary substantially in average memory, suggesting that implementation choices such as data structure selection and allocation patterns materially affect efficiency beyond correctness.

% 中文翻译：失败并非同质，平均通过样例数能够揭示解在何处发生崩溃。
\textbf{Finding 3. Failures are not homogeneous: average passed cases reveal the stage at which solutions break down.}
Average passed cases tends to remain high for frontier models across languages, indicating that many failures occur after passing a large portion of hidden tests and are consistent with late-stage corner cases. When response errors or severe compilation issues occur, average passed cases can drop and indicates early breakdowns that prevent deep test coverage. \texttt{deepseek-v3.2} illustrates this behavior in Java where average passed cases drops to 481.45, consistent with its unstable execution outcomes. This metric is complementary to cumulative runtime because it helps explain whether penalties arise from early pipeline failures or from solutions that are close to acceptance but miss a small set of cases.

In conclusion, efficiency differences in \projectname are shaped by penalty-accumulating runtime trajectories and by the conditional runtime, memory, and failure depth of accepted and near-miss solutions. This suggests actionable guidance for coding agents that goes beyond correctness. Agents should perform complexity-aware planning before coding, apply targeted revisions that replace high-cost approaches when time-limit signals appear, and explicitly trade off runtime and memory via data-structure choices. When failures occur, average passed cases help distinguish early breakdowns from late-stage corner-case errors, enabling more efficient refinement under fixed budgets.

\subsubsection{Comparison of Human-referenced Performance}
We evaluate human-referenced efficiency with ARB and AMB, which correspond to \emph{average runtime beats} and \emph{average memory beats} against the empirical distribution of human submissions for the same problem and programming language. In Figure~\ref{fig:inference-performance-comparison}, this dimension corresponds to the third row, which tracks cumulative averages of runtime beats and memory beats as tasks accumulate. The solid and dotted trajectories therefore indicate whether competitiveness against humans is stable over the task sequence, rather than being driven by a small subset of early and easy problems.

% 中文翻译：强模型在多语言上能够持续保持与人类相比具有竞争力的效率表现。
\textbf{Finding 1. Strong models sustain human-competitive efficiency across languages, and this competitiveness persists over the task sequence.}
Across languages, frontier models typically achieve ARB and AMB above 50, indicating that their accepted solutions are often faster and more memory efficient than a majority of human submissions. This pattern is most pronounced for \texttt{gpt-5.2}, which maintains consistently strong third-row trajectories and reaches particularly high values in Kotlin, with ARB 88.44 and AMB 83.12. \texttt{claude-4.5-opus}, \texttt{gemini-3-flash-preview}, and \texttt{gemini-3-pro-preview} also attain high beats in several languages, and their third-row curves remain relatively stable as more tasks accumulate. This suggests that human-competitive efficiency is not limited to a small number of easy problems, but is maintained across a broader slice of the benchmark.

% 中文翻译：人类对比指标将“已解决样本的条件效率”与“整体覆盖率”解耦。
\textbf{Finding 2. Human-referenced beats decouple conditional efficiency from benchmark coverage.}
Because ARB and AMB are defined only for accepted submissions, a model with low pass rate can still appear strong in beats if its few successful solutions are unusually efficient. For example, \texttt{deepseek-v3.2} attains a low pass rate of 11 in C++, yet its accepted submissions achieve ARB 73.11 and AMB 74.05. This selection effect is also visible in the third row, where some trajectories remain moderate even when the corresponding pass-rate curve is low. As a result, beats metrics quantify \emph{conditional} efficiency among solved cases, while pass rate measures \emph{coverage}, and the two should be interpreted jointly when comparing models.

% 中文翻译：运行时间 beats 与内存 beats 的差距揭示了相对人类解的不同效率取舍。
\textbf{Finding 3. The gap between runtime beats and memory beats reveals distinct tradeoffs relative to human solutions.}
In some settings, models achieve much higher ARB than AMB, indicating solutions that prioritize speed over memory compared to human submissions. For instance, in Python3 \texttt{gpt-5.2} attains ARB 68.06 but AMB 37.98, suggesting stronger relative runtime competitiveness than memory competitiveness. In other settings, ARB and AMB are closer, indicating more balanced efficiency. The solid and dotted third-row trajectories make these tradeoffs visible as tasks accumulate, complementing absolute runtime and memory statistics with an interpretable human-referenced scale.

In conclusion, ARB and AMB together with the third-row trajectories in Figure~\ref{fig:inference-performance-comparison} show that human-competitive efficiency can be sustained across tasks, yet it interacts with both benchmark coverage and runtime--memory tradeoffs. This suggests practical guidance for coding agents. Agents should optimize for beats only after securing correctness, choose a speed--memory tradeoff matching problem constraints, and report beats jointly with pass rate to avoid over-interpreting efficiency from few accepted cases.

\subsection{Evaluation for Self-Evolving Agents}

Self-evolving coding agents aim to improve solution quality and efficiency during a single inference process through iterative refinement. Existing code benchmarks typically evaluate only the final submission and therefore provide limited visibility into when and how such within-inference self-evolution helps. \projectname supports a more fine-grained evaluation because it exposes execution outcomes, partial progress signals, and efficiency statistics, enabling trajectory-level analyses in addition to endpoint scores. In what follows, we evaluate self-evolving agents from two complementary perspectives. We first summarize overall performance using the same standardized metrics as in the main results. We then study within-inference evolution trajectories across reflection and revision steps to quantify improvement dynamics.

\subsubsection{Self-Evolving Performance}

\begin{table}[htbp]
    \centering
    \footnotesize
    
    \caption{Results based on \texttt{gemini-3-flash-preview}. Vanilla and evolving agents are reported; \textbf{Imp.(\%)} denotes improvement(\%). Green indicates gains. Red denotes degradation.}
    \vspace{-0.3cm}
    \label{tab:self-evolving-gemini3-flash}
    \begingroup
    \setlength{\tabcolsep}{1.2pt}
    \renewcommand{\arraystretch}{0.4}
    \begin{tabular}{l c ccccccc ccccc}
    \toprule
    \textbf{Agent} & \multicolumn{8}{c}{\makecell{\textbf{Capability}\\\textbf{metrics}}} & \multicolumn{3}{c}{\makecell{\textbf{Efficiency}\\\textbf{metrics}}} & \multicolumn{2}{c}{\makecell{\textbf{Human}\\\textbf{metrics}}} \\
    \cmidrule(lr){2-9}\cmidrule(lr){10-12}\cmidrule(lr){13-14}
     & \textbf{PR} & \textbf{TLE} & \textbf{MLE} & \textbf{CE} & \textbf{RE} & \textbf{WA} & \textbf{TO} & \textbf{RpE} & \makecell{\textbf{AR}} & \makecell{\textbf{AM}} & \makecell{\textbf{APC}} & \makecell{\textbf{ARB}} & \makecell{\textbf{AMB}} \\
    
    \midrule
    \rowcolor{gray!15} \multicolumn{14}{c}{\textit{\textbf{Python3}}} \\
    vanilla  & 79 & 4 & 0 & 0 & 2 & 14 & 1 & 0 & 1376.19 & 56.59 & 750.89 & 73.28 & 36.62 \\
    evolving     & 87 & 3 & 0 & 0 & 1 & 9  & 0 & 0 & 1269.39 & 59.08 & 750.98 & 70.29 & 42.15 \\
    \textbf{Imp.(\%)} & \textcolor{impgreen}{10.1} & \textcolor{impgreen}{25.0} & 0 & 0 & \textcolor{impgreen}{50} & \textcolor{impgreen}{35.7} & \textcolor{impgreen}{100} & 0 & \textcolor{impgreen}{7.8} & \textcolor{impred}{4.4} & \textcolor{impred}{0.0} & \textcolor{impred}{4.1} & \textcolor{impgreen}{15.1} \\

    \midrule
    \rowcolor{gray!15} \multicolumn{14}{c}{\textit{\textbf{C++}}} \\
    vanilla      & 84 & 2 & 0 & 2 & 1 & 10 & 0 & 1 & 266.04  & 168.93 & 743.31 & 68.02 & 59.24 \\
    evolving & 99 & 0 & 0 & 0 & 0 & 1  & 0 & 0 & 142.60  & 148.43 & 749.86 & 88.99 & 73.14 \\
    \textbf{Imp.(\%)} & \textcolor{impgreen}{17.9} & \textcolor{impgreen}{100} & 0 & \textcolor{impgreen}{100} & \textcolor{impgreen}{100} & \textcolor{impgreen}{90} & 0 & \textcolor{impgreen}{100} & \textcolor{impgreen}{46.4} & \textcolor{impgreen}{12.1} & \textcolor{impred}{0.9} & \textcolor{impgreen}{30.8} & \textcolor{impgreen}{23.5} \\
    
    \midrule
    \rowcolor{gray!15} \multicolumn{14}{c}{\textit{\textbf{Java}}} \\
    vanilla    & 84 & 0 & 0 & 2 & 2 & 9  & 1 & 2 & 125.04  & 126.09 & 752.86 & 71.03 & 59.18 \\
    evolving  & 98 & 1 & 0 & 0 & 0 & 1  & 0 & 0 & 96.30   & 120.00 & 751.09 & 88.33 & 72.38 \\
    \textbf{Imp.(\%)} & \textcolor{impgreen}{16.7} & 0 & 0 & \textcolor{impgreen}{100} & \textcolor{impgreen}{100} & \textcolor{impgreen}{88.9} & \textcolor{impgreen}{100} & \textcolor{impgreen}{100} & \textcolor{impgreen}{23.0} & \textcolor{impgreen}{4.8} & \textcolor{impgreen}{0.2} & \textcolor{impgreen}{24.4} & \textcolor{impgreen}{22.3} \\

    \midrule
    \rowcolor{gray!15} \multicolumn{14}{c}{\textit{\textbf{Go}}} \\
    vanilla      & 82 & 1 & 0 & 9 & 0 & 7  & 0 & 1 & 139.22  & 22.01  & 739.46 & 76.22 & 63.48 \\
    evolving    & 95 & 0 & 0 & 0 & 0 & 5  & 0 & 0 & 111.64  & 18.35  & 754.17 & 81.52 & 67.94 \\
    \textbf{Imp.(\%)} & \textcolor{impgreen}{15.9} & \textcolor{impgreen}{100} & 0 & \textcolor{impgreen}{100} & 0 & \textcolor{impgreen}{28.6} & 0 & \textcolor{impgreen}{100} & \textcolor{impgreen}{19.8} & \textcolor{impgreen}{16.6} & \textcolor{impred}{2.0} & \textcolor{impgreen}{7.0} & \textcolor{impgreen}{7.0} \\
    
    \midrule
    \rowcolor{gray!15} \multicolumn{14}{c}{\textit{\textbf{Kotlin}}} \\
    vanilla  & 75 & 2 & 0 & 8 & 1 & 10 & 2 & 2 & 171.99  & 72.80  & 760.43 & 83.49 & 79.07 \\
    evolving & 95 & 1 & 0 & 0 & 0 & 4  & 0 & 0 & 122.83  & 77.88  & 749.38 & 83.58 & 67.21 \\
    \textbf{Imp.(\%)} & \textcolor{impgreen}{26.7} & \textcolor{impgreen}{50} & 0 & \textcolor{impgreen}{100} & \textcolor{impgreen}{100} & \textcolor{impgreen}{60} & \textcolor{impgreen}{100} & \textcolor{impgreen}{100} & \textcolor{impgreen}{28.6} & \textcolor{impred}{7.0} & \textcolor{impred}{1.5} & \textcolor{impgreen}{0.1} & \textcolor{impred}{15.0} \\
    \bottomrule
    \end{tabular}
    \endgroup
    \vspace{-0.7cm}
\end{table}

\begin{figure}[htbp]
    \vspace{-0.5cm}
    \centering
    \includegraphics[width=\linewidth]{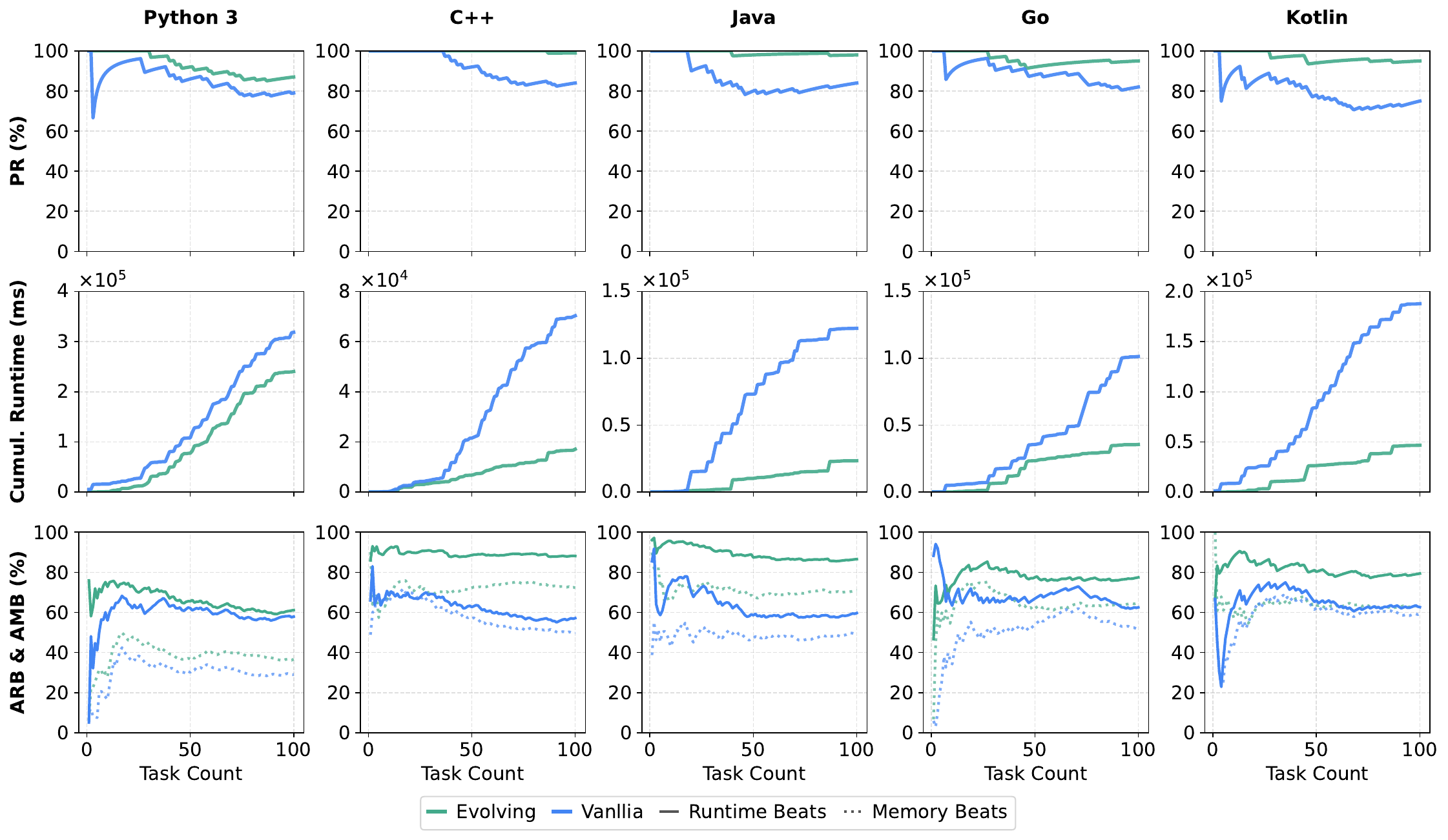}
    \vspace{-0.9cm}
    \caption{Performance comparison of evolving and vanilla agents within-inference.}
    \label{fig:evolution-performance-comparison}
    \vspace{-0.5cm}
\end{figure}

\leavevmode\\[-0.2em]\noindent
\textbf{Coding capability.}
Table~\ref{tab:self-evolving-gemini3-flash} compares a vanilla coding agent against an evolving variant for \texttt{gemini-3-flash-preview}. Across languages, the evolving agent yields relative pass-rate improvements of 10.1 to 26.7 percent. Figure~\ref{fig:evolution-performance-comparison} corroborates this in the first row by showing consistently higher pass-rate trajectories as tasks accumulate. The effect is strongest in compiled languages, where pass rates reach 99 in C++ and 98 in Java. These improvements are accompanied by broad reductions in execution blocking errors. Compilation error, runtime error, timeout, and response error frequently drop to zero, indicating that iterative refinement effectively repairs format and tooling issues. Beyond executability, the evolving agent also reduces algorithmic failures. Time limit exceeded and wrong answer counts decrease substantially, suggesting that self reflection helps identify inefficient approaches and missed corner cases. Overall, within inference refinement can convert many near miss submissions into accepted solutions under the same evaluation interface.

\textbf{Coding efficiency.}
Efficiency metrics show that the evolving agent improves runtime in every language, with average runtime decreasing by 7.8 percent in Python3 and by 19.8 to 46.4 percent in compiled languages. In Figure~\ref{fig:evolution-performance-comparison}, the second row confirms this trend, as the evolving agent accumulates substantially lower cumulative runtime and the gap widens as tasks accumulate. This pattern is consistent with reductions in time limit exceeded errors and suggests that evolution often replaces suboptimal algorithms with more efficient ones. Memory usage shows a mixed trend, decreasing in C++, Java, and Go but increasing modestly in Python3 and Kotlin, plausibly because the evolving agent introduces auxiliary data structures to ensure correctness or improve speed. Average passed cases changes only slightly, indicating that most gains come from turning partial solutions into fully accepted ones rather than uniformly increasing partial progress.

\textbf{Human-referenced comparison.}
Human-referenced metrics further suggest that evolution improves human calibrated efficiency in most languages. In Figure~\ref{fig:evolution-performance-comparison}, the third row shows that the evolving agent maintains higher runtime-beats and memory-beats trajectories than the vanilla agent in most settings, indicating sustained competitiveness against human submissions as tasks accumulate. Consistent with Table~\ref{tab:self-evolving-gemini3-flash}, runtime beats increase strongly in compiled languages, with ARB gains of 30.8 percent in C++ and 24.4 percent in Java, and smaller gains in Go and Kotlin. Memory beats increase in Python3, C++, Java, and Go, but decrease in Kotlin. This divergence mirrors the absolute memory trend and suggests that in long tail languages the evolving agent may trade memory for correctness or speed. Taken together, these results indicate that \projectname can quantify not only endpoint accuracy gains from evolution, but also whether those gains align with human efficiency distributions over the inference trajectory.

\subsubsection{Within-Inference Evolution Trajectories}

Figure~\ref{fig:evolution-performance-comparison} illustrates how \projectname enables trajectory-level evaluation of self-evolving coding agents. In the first row, the evolving agent maintains consistently higher pass-rate trajectories across languages, indicating that improvements persist as problems accumulate. In the second row, the evolving agent exhibits substantially lower cumulative runtime, with the widening gap suggesting compounding efficiency gains from fewer penalties and faster solutions. In the third row, the evolving agent sustains higher runtime-beats and memory-beats trajectories, showing that accepted solutions become more competitive relative to humans as inference progresses. Together, these signals complement endpoint metrics by revealing when and how within-inference refinement improves correctness and efficiency.

\section{Conclusion}

We introduced \projectname, a multilingual, execution based benchmark for evaluating LLM based coding systems under a unified online judge protocol. \projectname jointly measures error modes, resource efficiency, and human referential competitiveness, enabling comparisons that remain meaningful across high resource and long tail languages. Our results reveal systematic performance degradation in long tail languages, distinct failure modes driven by code level robustness and algorithmic inefficiency, and substantial variation in structured output reliability. We further demonstrate that \projectname supports within inference evaluation of self evolving agents by exposing improvement trajectories in pass rate, cumulative runtime, and beats metrics. We hope \projectname will facilitate more reliable evaluation and drive progress toward coding agents that are robust across languages, efficient under constraints, and competitive with human solutions.

\clearpage
%%
%% The next two lines define the bibliography style to be used, and
%% the bibliography file.
\bibliographystyle{ACM-Reference-Format}
\bibliography{main}

\onecolumn

% %% If your work has an appendix, this is the place to put it.
\appendix
\section{Code and Data Availability}
All code and data will be publicly available upon acceptance of this paper, including the benchmark query set, evaluation pipeline, backtest engine, and supplementary scripts for reproducing the reported results.

\section{Motivation}
\label{appx_sec:motivation}

The design of \projectname is driven by three complementary observations about the current evaluation landscape for LLM-based coding systems. In this section, we elaborate on each motivation in detail and explain how our benchmark addresses the corresponding gap.

\subsection{Evaluating Inference-Time Self-Evolution}

A growing body of work explores how LLM-based agents can improve during inference through iterative refinement, self-repair, and reflection~\citep{madaan2023selfrefine,shinn2023reflexion}. In the coding domain, these systems refine candidate solutions by analyzing execution feedback, diagnosing failure modes such as wrong answers or time-limit violations, and producing revised implementations within a single problem-solving episode. This inference-time self-evolution has been shown to substantially boost pass rates, sometimes surpassing the performance of stronger base models operating in a single-pass mode~\citep{olausson2024selfrepair}.

Despite the growing importance of such systems, existing code benchmarks are not designed to characterize this behavior. They evaluate only the final submitted solution and report a single correctness score, which collapses all intermediate dynamics into one number. As a result, they cannot answer questions such as: Does the system become more accurate as it iterates? Does efficiency improve alongside correctness, or does the system trade runtime for correctness? How many revision rounds are needed before gains saturate? These trajectory-level questions are critical for understanding self-evolving agents, yet they require benchmarks that expose per-iteration feedback signals, including execution outcomes, runtime, memory, and partial-progress indicators.

\projectname addresses this gap by recording the full sequence of submissions, execution verdicts, and resource statistics across reflection and revision steps. This enables trajectory-level analyses that separate one-shot generation quality from inference-time adaptation gains, quantify the speed and cost of improvement, and identify when self-evolution helps versus when it introduces regressions. Our empirical results confirm that this design reveals meaningful improvement dynamics. For example, the self-evolving agent achieves pass-rate improvements of 10 to 27 percent relative to the vanilla agent, and cumulative runtime decreases substantially as tasks accumulate, indicating compounding efficiency gains that would be invisible under endpoint-only evaluation.

\subsection{Calibrating Performance Against Human Ability}

Most code benchmarks report model performance in absolute units such as pass rate or accuracy, which are difficult to interpret without an external reference frame. A pass rate of 80 conveys limited information about whether the system is approaching, matching, or exceeding typical human performance on the same tasks. This interpretability gap is especially problematic as frontier models approach high accuracy on popular benchmarks, making it unclear whether remaining failures reflect fundamental limitations or merely reflect difficult edge cases that also challenge human programmers.

Human-referenced evaluation provides a complementary perspective by locating model performance within the empirical distribution of human submissions. On platforms such as LeetCode, every accepted submission receives percentile-based statistics that indicate the fraction of human solutions it outperforms in runtime and memory. These runtime beats and memory beats metrics answer a qualitatively different question from pass rate: not just whether the model solves the problem, but how its solution compares to the solutions produced by real human programmers on the same task and in the same programming language.

\projectname integrates these human-referenced signals as first-class evaluation metrics. By reporting average runtime beats and average memory beats alongside pass rate, our benchmark enables statements such as ``the model's accepted solutions outperform 70 percent of human submissions in runtime'' or ``the model achieves human-competitive memory efficiency in long-tail languages despite lower pass rates.'' These calibrated comparisons are essential for interpreting the practical significance of benchmark scores and for identifying settings where models are competitive with or superior to typical human programmers, which has implications for deployment decisions and trust in automated coding systems.

\subsection{Assessing Cross-Language Robustness Under Long-Tail Usage}

Existing code benchmarks are heavily concentrated on Python, which reflects its dominant position in both the LLM training data distribution and the developer ecosystem. According to the TIOBE Index for January 2026, Python accounts for 22.61 percent of market share, while long-tail languages such as Kotlin account for only 0.97 percent~\citep{tiobe2026}. This skew has two consequences for evaluation. First, benchmarks that evaluate only in Python cannot detect cross-language robustness issues, such as compilation errors in statically typed languages or API misuse in less-represented ecosystems. Second, the training data imbalance means that models are likely to have seen far fewer examples of correct Kotlin or Go code, which can translate into systematic performance degradation that single-language benchmarks would miss entirely.

Several prior efforts have extended evaluation to multiple languages~\citep{cassano2023multiple,ding2023crosscodeeval,raihan2025mhumaneval}, but they typically focus on translating existing Python benchmarks rather than designing evaluation protocols that explicitly target long-tail robustness. In particular, they rarely combine multilingual coverage with efficiency metrics or human-referenced comparisons, which limits their ability to diagnose whether performance gaps arise from algorithmic deficiencies, implementation-level friction, or language-specific toolchain issues.

\projectname includes five programming languages, spanning high-resource (Python3, C++, Java), moderate-resource (Go), and long-tail (Kotlin) settings, all evaluated under a unified online judge protocol with identical metrics. This design enables systematic analysis of how pass rate, error profiles, runtime, memory, and human-referenced beats shift as language resource availability decreases. Our results confirm that this multilingual design exposes robustness gaps that are invisible in Python-only evaluation: compilation errors increase sharply in long-tail languages, overall pass rates degrade systematically, and the gap between high-resource and long-tail performance varies meaningfully across models. These findings motivate the inclusion of long-tail languages as a standard component of code benchmark design.

\section{Case Studies}
\label{appx_sec:case-studies}

\subsection{Agent Construction Details}
\label{appx_sec:agent-construction}

In the main paper, we introduced two standardized evaluation interfaces: a lightweight \textbf{coding agent} and a \textbf{self-evolving coding agent}. This subsection provides additional implementation details, including prompt templates, output parsing rules, the submission pipeline, and the reflection and revision procedure.

\subsubsection{Coding Agent}

The coding agent is designed to be minimally augmented so that observed performance differences reflect the underlying model's capabilities rather than agent-level engineering. Its construction follows four steps.

\textbf{Step 1: Prompt Construction.}
For each problem and target programming language, the agent assembles a standardized prompt that consists of three components: (1) the full natural-language problem statement, including constraints and input--output examples; (2) any embedded images associated with the problem, resolved to local assets; and (3) the language-specific starter code template provided by the platform. These components are concatenated into a fixed-format input string. The prompt explicitly instructs the model to return a structured response containing two fields: a \texttt{reasoning} field for intermediate analysis and a \texttt{code} field for the final solution. This format constraint ensures deterministic parsing regardless of the model's default output style. An illustrative prompt is shown in Appendix Figure~\ref{appx_fig:case-add-two-numbers}.

\textbf{Step 2: Model Inference.}
The assembled prompt is sent to the target LLM with reasoning enabled by default and a maximum output token budget of 16,384 tokens. No system-level instructions, few-shot examples, or retrieval augmentation are provided. The model generates its response in a single forward pass without access to external tools, memory, or execution feedback.

\textbf{Step 3: Deterministic Code Extraction.}
The agent parses the model's response and extracts the content of the \texttt{code} field using a deterministic extraction rule. If the response does not conform to the expected format, the submission is recorded as a response error. This strict parsing ensures that evaluation failures due to format non-compliance are explicitly captured rather than silently corrected.

\textbf{Step 4: Online Judge Submission and Feedback Collection.}
The extracted code is submitted to the LeetCode online judge for execution-based evaluation against hidden test cases. The judge returns a verdict (Accepted, Wrong Answer, Time Limit Exceeded, Memory Limit Exceeded, Compile Error, or Runtime Error), the number of passed test cases, and, for accepted submissions, runtime in milliseconds, memory usage in megabytes, and percentile-based runtime beats and memory beats statistics computed against the distribution of human submissions. All signals are recorded for downstream analysis.

\subsubsection{Self-Evolving Coding Agent}

The self-evolving coding agent extends the coding agent with a controlled, within-inference reflection and revision loop. Its construction adds three additional steps after the initial submission.

\textbf{Step 5: Execution Feedback Analysis.}
After the initial submission (Steps 1--4), the agent receives the judge's verdict and diagnostic information. If the submission is accepted, no further refinement is performed. If the submission fails, the agent constructs a reflection prompt that includes: (a) the original problem statement, (b) the previously submitted code, (c) the judge's verdict and error details (e.g., failed test case input and expected output for wrong answers, or error messages for compile and runtime errors), and (d) an explicit instruction to analyze the failure and propose targeted improvements.

\textbf{Step 6: Reflection and Revision.}
The model receives the reflection prompt and generates an analysis of the failure, identifying potential root causes such as unhandled corner cases, incorrect algorithmic logic, suboptimal time or space complexity, or language-specific syntax issues. Based on this analysis, the model produces a revised solution. The revised code is extracted using the same deterministic parsing rule as in Step 3 and resubmitted to the online judge.

\textbf{Step 7: Iteration and Termination.}
Steps 5 and 6 are repeated for up to three reflection and revision rounds (configurable). The loop terminates early if any revised submission achieves an Accepted verdict. At each iteration, the full execution feedback is recorded, enabling trajectory-level analysis of how correctness, runtime, memory, and human-referenced beats evolve across revision steps.

\textbf{Key Design Constraints.}
Throughout the self-evolution process, only the solution code is modified. The problem specification, prompt schema, evaluation interface, and model parameters remain fixed. This ensures that observed improvements are attributable solely to inference-time refinement rather than changes in task formulation, prompt engineering, or external augmentation. By constraining refinement to a fixed revision budget within a single inference episode, the agent enables reproducible analysis of improvement dynamics.

\subsubsection{Prompt Example}

The following case study illustrates the prompt format used by the coding agent. The input includes the problem statement, examples with visual illustrations, constraints, and the language-specific starter template.

\label{appx_sec:case-add-two-numbers}
\begin{CaseStudyBox}
    \textbf{Problem.} Add Two Numbers \\
    \textbf{Description:} You are given two non-empty linked lists representing two non-negative integers. The digits are stored in reverse order, and each of their nodes contains a single digit. Add the two numbers and return the sum as a linked list.\\

    You may assume the two numbers do not contain any leading zero, except the number 0 itself. \\
    \textbf{Example 1:}
    % Note: floats (figure/table) cannot appear reliably inside tcolorbox.
    \begin{center}
        \includegraphics[width=0.25\linewidth]{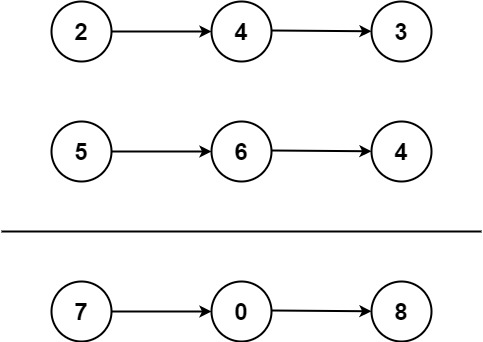}
    \end{center}
    Input: l1 = [2,4,3], l2 = [5,6,4] \\
    Output: [7,0,8] \\
    Explanation: 342 + 465 = 807. \\
    \textbf{Example 2:}
    Input: l1 = [0], l2 = [0] \\
    Output: [0] \\
    \textbf{Example 3:}
    Input: l1 = [9,9,9,9,9,9,9], l2 = [9,9,9,9] \\
    Output: [8,9,9,9,0,0,0,1] \\
    \textbf{Constraints:}
    1. The number of nodes in each linked list is in the range [1, 100]. \\
    2. 0 <= Node.val <= 9 \\
    3. It is guaranteed that the list represents a number that does not have leading zeros. \\
    \textbf{Template:}
\begin{lstlisting}[style=evocode,language=Python]
# Definition for singly-linked list.
# class ListNode:
#     def __init__(self, val=0, next=None):
#         self.val = val
#         self.next = next
class Solution:
    def addTwoNumbers(
        self,
        l1: Optional[ListNode],
        l2: Optional[ListNode],
    ) -> Optional[ListNode]:
\end{lstlisting}
\end{CaseStudyBox}
\vspace{-0.5cm}
\captionof{figure}{Prompt example for coding agent.}
\label{appx_fig:case-add-two-numbers}

\subsubsection{Self-Evolving Coding Agent Prompts}
\label{appx_sec:self-evolving-prompts}

The self-evolving coding agent uses two types of reflection prompts depending on the judge verdict: a \emph{bugfix} reflection when the submission fails and an \emph{optimization} reflection when the submission is accepted but performance is below a configurable threshold. The agent expects a structured JSON response with three fields: \texttt{analysis}, \texttt{improved\_reasoning}, and \texttt{improved\_code}. The reflection loop runs for at most three rounds and terminates early when the solution is accepted with competitive performance (both runtime beats and memory beats exceeding 65\%). Throughout the process, only the solution code is modified; the problem specification, prompt schema, and model parameters remain fixed.

\begin{CaseStudyBox}
\textbf{Output Schema (Coding Agent):}
\begin{lstlisting}[style=evocode,language=Python]
class CodeResponse:
    reasoning: str  # The reasoning process
    result: str     # The generated code
\end{lstlisting}

\textbf{Output Schema (Self-Evolving Agent Reflection):}
\begin{lstlisting}[style=evocode,language=Python]
class ReflectionResponse:
    analysis: str           # Why the previous solution failed
                            # or how it can be optimized
    improved_reasoning: str # Improved reasoning process
    improved_code: str      # Improved solution code
\end{lstlisting}

\vspace{0.3em}
\textbf{Bugfix Reflection --- System Prompt:}
\begin{lstlisting}[style=evocode]
You are an expert programmer debugging a LeetCode solution.
You need to carefully analyze why it failed and provide a
fixed solution.

IMPORTANT DEBUGGING GUIDELINES:
- Check boundary conditions and edge cases
- Verify your algorithm logic step by step
- Consider special inputs like empty arrays, single elements,
  negative numbers, etc.
- Make sure you understand the problem requirements correctly

You MUST:
1. Carefully analyze the error type and evaluation result
2. Identify the root cause of the failure
3. Provide a completely corrected solution. Do not assume the
   original approach is correct -- validate the algorithm
   itself and fix or replace it if necessary
4. Make sure your code is syntactically correct and handles
   all edge cases

Output format: JSON with fields:
  "analysis", "improved_reasoning", "improved_code"
\end{lstlisting}

\textbf{Bugfix Reflection --- User Prompt:}
\begin{lstlisting}[style=evocode]
## Problem
{problem_statement}

## Your Previous Solution (Round {round_num - 1})
### Reasoning:
{previous_reasoning}
### Code:
{previous_code}

## Evaluation Result
Status: {status}
Passed Cases: {passed}/{total}
Details: {error_details}

## Task
Your solution got "{status}". Analyze why it failed and
provide a fixed solution.
\end{lstlisting}

\vspace{0.3em}
\textbf{Optimization Reflection --- System Prompt:}
\begin{lstlisting}[style=evocode]
You are an expert programmer optimizing a LeetCode solution.
Your previous solution was ACCEPTED but the performance can
be improved:
- Runtime: {runtime}ms (beats {runtime_beats}%)
- Memory: {memory}MB (beats {memory_beats}%)

OPTIMIZATION STRATEGIES TO CONSIDER:
1. Time: more efficient algorithms, better data structures,
   reduce unnecessary computations, DP or memoization
2. Space: in-place algorithms, avoid unnecessary copies,
   iterative vs recursive, reuse variables

IMPORTANT:
- The optimized solution MUST still be correct
- Focus on reducing time complexity first, then space
- Explain the complexity improvement in your analysis

Output format: JSON with fields:
  "analysis", "improved_reasoning", "improved_code"
\end{lstlisting}

\textbf{Optimization Reflection --- User Prompt:}
\begin{lstlisting}[style=evocode]
## Problem
{problem_statement}

## Your Previous Solution (Round {round_num - 1}) - ACCEPTED
### Reasoning:
{previous_reasoning}
### Code:
{previous_code}

## Current Performance
Status: Accepted
Runtime: {runtime}ms (beats {runtime_beats}%)
Memory: {memory}MB (beats {memory_beats}%)

## Task
Your solution is correct but can be optimized:
1. Maintains correctness (passes all test cases)
2. Has better time complexity if possible
3. Has better space complexity if possible
\end{lstlisting}

\vspace{0.3em}
\textbf{Termination Criteria:}\\
The reflection loop terminates when: (1) the maximum number of rounds (default 3) is reached; (2) the solution achieves Accepted with both runtime beats and memory beats exceeding 65\%; or (3) the model fails to produce a valid reflection response. The agent always retains the best submission, preferring Accepted solutions and, among those, the one with the highest runtime beats.
\end{CaseStudyBox}
\vspace{-0.5cm}
\captionof{figure}{Complete prompt templates and output schemas for the self-evolving coding agent, including the bugfix reflection prompt (for failed submissions), the optimization reflection prompt (for accepted but suboptimal submissions), and termination criteria.}
\label{appx_fig:self-evolving-prompts}

\subsection{Accepted Test Cases}
In this subsection, we present a representative selection of test cases where the model successfully generated correct solutions, achieving an ``Accepted'' status. These examples demonstrate the agent's proficiency in handling a diverse range of data structures and algorithmic paradigms, specifically focusing on \textbf{arrays}, \textbf{math}, \textbf{dynamic programming} and \textbf{graph theory}. Through these cases, we illustrate the model's capability to interpret complex logic, implement efficient algorithms (such as monotonic stacks and bitwise operations), and satisfy strict execution constraints.
\subsubsection{Case 1: Steps to Make Array Non-decreasing}

In Case 1, the model addresses the problem ``Steps to Make Array Non-decreasing,'' a task that requires sophisticated algorithmic reasoning. A naive simulation of the removal process would result in a time complexity of $O(N^2)$, which violates the given constraints ($N=10^5$). The model successfully identifies the optimal approach using a \textbf{monotonic stack}, demonstrating the ability to transform a simulation problem into a linear-time $O(N)$ solution.

The generated explanation reveals a clear chain of thought: the model correctly deduces that an element's removal time depends on the maximum removal time of the smaller elements it covers. By maintaining a stack of $(value, time)$ tuples, the code effectively calculates the "collision" logic required to determine when each element is removed. The evaluation results confirm the solution's correctness and efficiency, passing all 87 test cases and outperforming 82.5\% of Python3 submissions in runtime. This case highlights the model's capacity to handle complex state transitions and data structure optimizations without human intervention.
% 需要分析这个样例，有一些分析才行

\label{appx_sec:case-Steps-to-Make-Array-Non-decreasing}
\begin{CaseStudyBox}
    \textbf{Problem.} Steps to Make Array Non-decreasing \\
    \textbf{Tags}: 
Array,
Dynamic Programming,
Bit Manipulation\\
    \textbf{Description:} You are given a 0-indexed integer array nums. In one step, remove all elements nums[i] where nums[i - 1] > nums[i] for all 0 < i < nums.length.\\

Return the number of steps performed until nums becomes a non-decreasing array\\
    \textbf{Example 1:}\\
    % Note: floats (figure/table) cannot appear reliably inside tcolorbox.
    Input: nums = [5,3,4,4,7,3,6,11,8,5,11] \\
    Output: 3 \\
    Explanation: The following are the steps performed:\\
    Step 1: [5,3,4,4,7,3,6,11,8,5,11] becomes [5,4,4,7,6,11,11]\\
    Step 2: [5,4,4,7,6,11,11] becomes [5,4,7,11,11]\\
    Step 3: [5,4,7,11,11] becomes [5,7,11,11]
    
    [5,7,11,11] is a non-decreasing array. Therefore, we return 3. \\
    \textbf{Example 2:}\\
    Input: nums = [4,5,7,7,13] \\
    Output: 0 \\
    Explanation: nums is already a non-decreasing array. Therefore, we return 0.\\
    \textbf{Constraints:}\\
    $\bullet$ 1 <= nums.length <= $10^5$\\
    $\bullet$ 1 <= nums[i] <= $10^9$\\
    \textbf{Template:}
\begin{lstlisting}[style=evocode,language=Python]
class Solution:
    def totalSteps(self, nums: List[int]) -> int:
\end{lstlisting}
\textbf{Model inference result:}
\begin{lstlisting}[style=evocode,language=Python]
The problem asks for the number of steps to make an array non-decreasing by removing elements `nums[i]` where `nums[i-1] > nums[i]` in each step. This can be solved using a monotonic stack approach to track the 'time' each element takes to be removed.

For each element `x` in `nums`, we want to find the nearest element to its left that is strictly greater than `x` (let's call it the 'killer'). However, the 'killer' might be busy removing other elements that are between it and `x`. 

We maintain a stack of pairs `(value, time)`, where `time` is the step at which the `value` is removed. When we encounter a new element `x`:
1. We pop elements from the stack that are less than or equal to `x`. These elements cannot remove `x`. 
2. As we pop them, we keep track of the maximum removal time `t` among them. This `t` represents the number of steps required to clear out the elements between the potential 'killer' (the current top of the stack) and `x`.
3. If the stack is empty after popping, it means there is no element to the left of `x` that is greater than `x`, so `x` will never be removed (`time = 0`).
4. If the stack is not empty, the element at the top is the 'killer' for `x`. `x` will be removed at step `t + 1`. We then push `(x, t + 1)` onto the stack.
5. The answer is the maximum removal time recorded across all elements.

This approach ensures that each element is pushed and popped from the stack at most once, resulting in an O(n) time complexity and O(n) space complexity.

#
# @lc app=leetcode id=2289 lang=python3
#
# [2289] Steps to Make Array Non-decreasing
#

# @lc code=start
class Solution:
    def totalSteps(self, nums: List[int]) -> int:
        # stack will store pairs of (value, time_to_be_removed)
        stack = []
        max_steps = 0
        
        for x in nums:
            t = 0
            # Pop elements that are smaller than or equal to current element x.
            # These elements are either already removed or will be 'blocked' by x.
            # We need to find the max removal time among elements that x 'overtakes'.
            while stack and stack[-1][0] <= x:
                t = max(t, stack.pop()[1])
            
            if not stack:
                # No element to the left is greater than x, so x stays forever.
                stack.append((x, 0))
            else:
                # The element at stack[-1] is greater than x and will eventually remove it.
                # However, it must first remove all elements that were processed 
                # between it and x (which took 't' steps).
                t += 1
                stack.append((x, t))
                max_steps = max(max_steps, t)
                
        return max_steps
# @lc code=end
\end{lstlisting}
\textbf{Evaluation result:}
\textcolor{impgreen}{Accepted}\\
87/87 cases passed (105 ms)\\
Your runtime beats 82.5 \% of python3 submissions\\
Your memory usage beats 51 \% of python3 submissions (33 MB)
\end{CaseStudyBox}
\vspace{-0.5cm}
\captionof{figure}{Case study of the ``Steps to Make Array Non-decreasing'' problem. The figure illustrates the complete generation pipeline, including the problem description, the model's derivation of the monotonic stack approach, the generated solution, and the successful evaluation metrics.}
\label{appx_fig:case-Steps-to-Make-Array-Non-decreasing}
\subsubsection{Case 2: Number of Effective Subsequences}
\noindent
In Case 2, the model tackles ``Number of Effective Subsequences,'' a problem that demands a deep understanding of combinatorial mathematics and bitwise operations. A brute-force approach iterating through all subsequences would yield a time complexity of $O(2^N)$, which is computationally infeasible. The model successfully identifies that this is a counting problem best solved using the \textbf{Principle of Inclusion-Exclusion (PIE)} combined with \textbf{Sum Over Subsets (SOS) Dynamic Programming}.

The model's inference trace demonstrates a high level of abstraction: it correctly formulates the problem as finding the number of subsets that maintain the full bitwise OR sum, and then subtracting this from the total. It employs an optimization technique—mapping the active bits of the total OR sum to a compact range—to minimize the SOS DP state space to $O(d \cdot 2^d)$, where $d$ is the number of set bits (at most 20). The evaluation results (1041/1041 cases passed) and the high efficiency ranking (beating 71.26\% of submissions) validate the model's ability to implement complex mathematical algorithms correctly.
\label{appx_sec:case-Number-of-Effective-Subsequences}
\begin{CaseStudyBox}
    \textbf{Problem.} Number of Effective Subsequences \\
    \textbf{Tags}: 
Array,
Math,
Dynamic Programming
Bit Manipulation\\
    \textbf{Description:} You are given an integer array nums.The strength of the array is defined as the bitwise OR of all its elements.A subsequence is considered effective if removing that subsequence strictly decreases the strength of the remaining elements.Return the number of effective subsequences in nums. Since the answer may be large, return it modulo $10^9$ + 7.\\

The bitwise OR of an empty array is 0.\\
    \textbf{Example 1:}\\
    % Note: floats (figure/table) cannot appear reliably inside tcolorbox.
    Input: nums = [1,2,3] \\
    Output: 3 \\
    Explanation: 
    
    The Bitwise OR of the array is 1 OR 2 OR 3 = 3.
Subsequences that are effective are:

[1, 3]: The remaining element [2] has a Bitwise OR of 2.

[2, 3]: The remaining element [1] has a Bitwise OR of 1.

[1, 2, 3]: The remaining elements [] have a Bitwise OR of 0.\\
Thus, the total number of effective subsequences is 3.\\
    \textbf{Example 2:}\\
    Input: nums = [7,4,6] \\
    Output: 4 \\
    Explanation: 
    
    The Bitwise OR of the array is 7 OR 4 OR 6 = 7.
    Subsequences that are effective are:

    [7]: The remaining elements [4, 6] have a Bitwise OR of 6.

    [7, 4]: The remaining element [6] has a Bitwise OR of 6.

    [7, 6]: The remaining element [4] has a Bitwise OR of 4.

    [7, 4, 6]: The remaining elements [] have a Bitwise OR of 0.

    Thus, the total number of effective subsequences is 4.\\
    \textbf{Example 3:}\\
    Input: nums = [8,8] \\
    Output: 1 \\
    Explanation: 
    
    The Bitwise OR of the array is 8 OR 8 = 8.
    
    Only the subsequence [8, 8] is effective since removing it leaves [] which has a Bitwise OR of 0.
    
    Thus, the total number of effective subsequences is 1.\\
    \textbf{Constraints:}\\
    $\bullet$ 1 <= nums.length <= $10^5$\\
    $\bullet$ 1 <= nums[i] <= $10^6$\\
    \textbf{Template:}
\begin{lstlisting}[style=evocode,language=Python]
class Solution:
    def countEffective(self, nums: List[int]) -> int:
\end{lstlisting}
\textbf{Model inference result:}
\begin{lstlisting}[style=evocode,language=Python]
1.  The strength of an array is the bitwise OR of its elements. Let $S$ be the OR of all elements in `nums`.
2.  A subsequence is effective if removing it strictly decreases the strength of the remaining elements. Let $B$ be the subsequence (set of indices) and $R = A \setminus B$ be the remaining elements. The condition is $OR_{i \in R} (nums[i]) < S$.
3.  Since every $nums[i]$ is a submask of $S$, the OR of any subset $R$ is a submask of $S$. Thus, $OR_{i \in R} (nums[i]) < S$ is equivalent to $OR_{i \in R} (nums[i]) \neq S$.
4.  The number of effective subsequences is $2^n - f(S)$, where $f(S)$ is the number of subsets $R \subseteq nums$ such that $OR_{i \in R} (nums[i]) = S$.
5.  By the Principle of Inclusion-Exclusion (or Inverse Fast Walsh-Hadamard Transform), $f(S) = \sum_{m \subseteq S} (-1)^{\text{popcount}(S) - \text{popcount}(m)} 2^{c(m)}$, where $c(m)$ is the count of $i$ such that $nums[i]$ is a submask of $m$.
6.  The result is $2^n - f(S) = \sum_{m \subset S} (-1)^{\text{popcount}(S) - \text{popcount}(m) - 1} 2^{c(m)} \pmod{10^9+7}$.
7.  We compute $c(m)$ for all submasks $m$ of $S$ using SOS DP. To handle the case where $S$ has non-contiguous bits, we map the bits of $S$ to a compact range $[0, \text{popcount}(S)-1]$.
8.  The complexity is $O(n + d 2^d)$ where $d = \text{popcount}(S) \le 20$. This fits within the time limits.

#
# @lc app=leetcode id=3757 lang=python3
#
# [3757] Number of Effective Subsequences
#

# @lc code=start
class Solution:
    def countEffective(self, nums: List[int]) -> int:
        MOD = 10**9 + 7
        total_or = 0
        for x in nums:
            total_or |= x
        
        if total_or == 0:
            return 0
            
        bits = []
        for i in range(21):
            if (total_or >> i) & 1:
                bits.append(i)
        
        d = len(bits)
        bit_map = {b: i for i, b in enumerate(bits)}
        
        freq = [0] * (1 << d)
        for x in nums:
            compact_x = 0
            for i, b in enumerate(bits):
                if (x >> b) & 1:
                    compact_x |= (1 << i)
            freq[compact_x] += 1
            
        # SOS DP to find c(m): number of elements that are submasks of m
        c = freq
        for i in range(d):
            bit = 1 << i
            for mask in range(1 << d):
                if mask & bit:
                    c[mask] += c[mask ^ bit]
        
        # Precompute powers of 2
        pow2 = [1] * (len(nums) + 1)
        for i in range(1, len(nums) + 1):
            pow2[i] = (pow2[i-1] * 2) % MOD
            
        # Precompute popcounts
        popcounts = [0] * (1 << d)
        for i in range(1, 1 << d):
            popcounts[i] = popcounts[i >> 1] + (i & 1)
            
        # Result = sum_{m < S} (-1)^(popcount(S) - popcount(m) - 1) * 2^c(m)
        ans = 0
        for m_idx in range(1 << d):
            if m_idx == (1 << d) - 1:
                continue
            
            exponent = d - popcounts[m_idx] - 1
            term = pow2[c[m_idx]]
            if exponent % 2 == 1:
                ans = (ans - term) % MOD
            else:
                ans = (ans + term) % MOD
                
        return ans % MOD
# @lc code=end
\end{lstlisting}
\textbf{Evaluation result:}
\textcolor{impgreen}{Accepted}

1041/1041 cases passed (4820 ms)

Your runtime beats 71.26 \% of python3 submissions

Your memory usage beats 33.33 \% of python3 submissions (80.8 MB)
\end{CaseStudyBox}
\vspace{-0.5cm}
\captionof{figure}{Case study of the ``Number of Effective Subsequences'' problem. The model demonstrates advanced mathematical reasoning by applying the Principle of Inclusion-Exclusion (PIE) and Sum Over Subsets (SOS) DP to optimize the solution complexity.}
\label{appx_fig:case-Number-of-Effective-Subsequences}

\subsubsection{Case 3: Remove Methods From Project}
\noindent
In Case 3, the model solves ``Remove Methods From Project,'' a graph theory problem involving reachability analysis and constraint verification. The solution requires a two-step logic: first, identifying the connected component of ``suspicious'' nodes starting from node $k$, and second, validating that no external nodes invoke any node within this component.

The model correctly decomposes the problem, employing a standard Breadth-First Search (BFS) to mark all reachable nodes from the source $k$. Crucially, it recognizes the necessity of a post-traversal validation step: iterating through the edge list to detect if any ``safe'' node directs to a ``suspicious'' node, which is the failure condition for removal. The implementation is highly optimized, using an adjacency list for traversal and a boolean array for state tracking, resulting in linear time complexity $O(V + E)$. The evaluation confirms the efficiency of this approach, with the generated code outperforming 99.38\% of Python3 submissions.
\label{appx_sec:case-Remove-Methods-From-Project}
\begin{CaseStudyBox}
    \textbf{Problem.} Remove Methods From Project \\
    \textbf{Tags}: 
Depth-First Search,
Breadth-First Search,
Graph Theory\\
    \textbf{Description:} You are maintaining a project that has n methods numbered from 0 to n - 1.

You are given two integers n and k, and a 2D integer array invocations, where invocations[i] = [$a_i$, $b_i$] indicates that method $a_i$ invokes method $b_i$.

There is a known bug in method k. Method k, along with any method invoked by it, either directly or indirectly, are considered suspicious and we aim to remove them.

A group of methods can only be removed if no method outside the group invokes any methods within it.\\

Return an array containing all the remaining methods after removing all the suspicious methods. You may return the answer in any order. If it is not possible to remove all the suspicious methods, none should be removed.\\
    \textbf{Example 1:}\\
    % Note: floats (figure/table) cannot appear reliably inside tcolorbox.
    Input: n = 4, k = 1, invocations = [[1,2],[0,1],[3,2]] \\
    Output: [0,1,2,3] \\
    Explanation: 
    \begin{center}
        \includegraphics[width=0.25\linewidth]{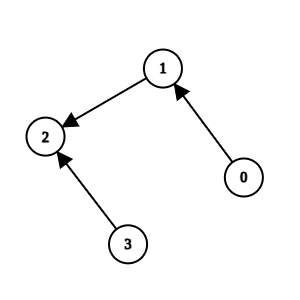}
    \end{center}
    Method 2 and method 1 are suspicious, but they are directly invoked by methods 3 and 0, which are not suspicious. We return all elements without removing anything.\\
    \textbf{Example 2:}\\
    Input: n = 5, k = 0, invocations = [[1,2],[0,2],[0,1],[3,4]]\\
    Output: [3,4] \\
    Explanation: 
    \begin{center}
        \includegraphics[width=0.25\linewidth]{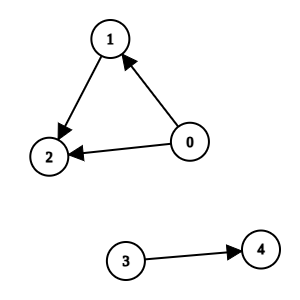}
    \end{center}
    Methods 0, 1, and 2 are suspicious and they are not directly invoked by any other method. We can remove them.\\
    \textbf{Example 3:}\\
    Input: n = 3, k = 2, invocations = [[1,2],[0,1],[2,0]]\\
    Output: [ ] \\
    Explanation: 
    \begin{center}
        \includegraphics[width=0.25\linewidth]{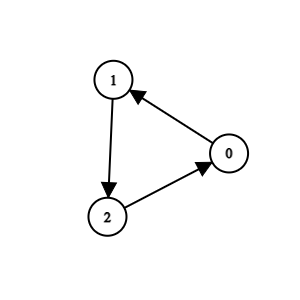}
    \end{center}
    All methods are suspicious. We can remove them.\\
    \textbf{Constraints:}\\
    $\bullet$ 1 <= n <= $10^5$\\
    $\bullet$ 0 <= k <= n-1\\
    $\bullet$ 0 <= invocations.length <= 2 * $10^5$\\
    $\bullet$ invocations[i] == [$a_i$, $b_i$]\\
    $\bullet$ 0 <= $a_i$, $b_i$ <= n-1\\
    $\bullet$ $a_i$ != $b_i$\\
    $\bullet$ invocations[i] != invocations[j]\\
    \textbf{Template:}
\begin{lstlisting}[style=evocode,language=Python]
class Solution:
    def remainingMethods(self, n: int, k: int, invocations: List[List[int]]) -> List[int]:
\end{lstlisting}
\textbf{Model inference result:}
\begin{lstlisting}[style=evocode,language=Python]
1. **Identify Suspicious Methods**: A method is suspicious if it is method `k` or if it is reachable from method `k` in the invocation graph. We can use Breadth-First Search (BFS) starting from `k` to find all reachable nodes.
2. **Check Removal Condition**: The set of suspicious methods can only be removed if no method outside this set calls any method inside this set. We iterate through all the edges `[u, v]` in the `invocations` list. If `u` is not suspicious but `v` is suspicious, the condition is violated, and we cannot remove the methods.
3. **Return Results**: If the removal condition is met, return all method IDs that are not suspicious. Otherwise, return all method IDs from `0` to `n - 1`.
4. **Complexity**: Building the adjacency list takes O(E) time. BFS takes O(V + E) time. Checking the removal condition takes O(E) time. Constructing the result list takes O(V) time. The total time complexity is O(V + E) and space complexity is O(V + E), where V = n and E = invocations.length.

#
# @lc app=leetcode id=3310 lang=python3
#
# [3310] Remove Methods From Project
#

# @lc code=start
from collections import deque
from typing import List

class Solution:
    def remainingMethods(self, n: int, k: int, invocations: List[List[int]]) -> List[int]:
        # 1. Build adjacency list for the invocation graph
        adj = [[] for _ in range(n)]
        for u, v in invocations:
            adj[u].append(v)
        
        # 2. Identify all suspicious methods using BFS starting from k
        suspicious = [False] * n
        suspicious[k] = True
        queue = deque([k])
        
        while queue:
            u = queue.popleft()
            for v in adj[u]:
                if not suspicious[v]:
                    suspicious[v] = True
                    queue.append(v)
        
        # 3. Check if any non-suspicious method invokes a suspicious one
        can_remove = True
        for u, v in invocations:
            if not suspicious[u] and suspicious[v]:
                can_remove = False
                break
        
        # 4. Return the result based on the check
        if can_remove:
            # Return only non-suspicious methods
            return [i for i in range(n) if not suspicious[i]]
        else:
            # Return all methods
            return [i for i in range(n)]
# @lc code=end
\end{lstlisting}
\textbf{Evaluation result:}
\textcolor{impgreen}{Accepted}\\
775/775 cases passed (191 ms)\\
Your runtime beats 99.38 \% of python3 submissions\\
Your memory usage beats 87.58 \% of python3 submissions (99.9 MB)
\end{CaseStudyBox}
\vspace{-0.5cm}
\captionof{figure}{Case study of the ``Remove Methods From Project'' problem. The model demonstrates effective graph reasoning by combining BFS for reachability analysis with an edge-based constraint validation step.}
\label{appx_fig:case-Remove-Methods-From-Project}

\subsection{Compile Error Resolution via Reflection Agent}
\noindent
This case exemplifies the capability of the \textbf{Reflection Agent} to handle language-specific syntax errors that prevent code execution. The model initially formulated a correct algorithmic strategy using Dijkstra's algorithm to solve the ``Find Minimum Time to Reach Last Room I'' problem. However, the submission failed with a \textbf{Compile Error} due to an ambiguous import of \texttt{PriorityQueue} in the Kotlin environment.

Upon receiving the error signal, the Reflection Agent did not merely regenerate the code stochastically. Instead, it performed a targeted diagnosis: it identified the conflict between the explicit import and the environment's implicit imports. Consequently, the agent refined the code by adopting a wildcard import (\texttt{java.util.*}) and simplifying the data structure from a custom data class to a primitive \texttt{IntArray} to reduce overhead. This iterative self-correction process allowed the system to recover from a fatal syntax error and achieve a fully accepted solution with competitive runtime performance.
\label{appx_sec:case-Find-Minimum-Time-to-Reach-Last-Room-I}
\begin{CaseStudyBox}
    \textbf{Problem.} Find Minimum Time to Reach Last Room I \\
    \textbf{Tags:} Graph Theory,
Heap (Priority Queue),
Matrix,
Shortest Path\\
    \textbf{Description:} There is a dungeon with n x m rooms arranged as a grid.\\

You are given a 2D array moveTime of size n x m, where moveTime[i][j] represents the minimum time in seconds after which the room opens and can be moved to. You start from the room (0, 0) at time t = 0 and can move to an adjacent room. Moving between adjacent rooms takes exactly one second.\\

Return the minimum time to reach the room (n - 1, m - 1).\\

Two rooms are adjacent if they share a common wall, either horizontally or vertically.\\
    \textbf{Example 1:}\\
    % Note: floats (figure/table) cannot appear reliably inside tcolorbox.
    Input: moveTime = [[0,4],[4,4]] \\
    Output: 6 \\
    Explanation: 
    The minimum time required is 6 seconds.

    $\bullet$ At time t == 4, move from room (0, 0) to room (1, 0) in one second.\\
    $\bullet$ At time t == 5, move from room (1, 0) to room (1, 1) in one second.\\
    \textbf{Example 2:}\\
    Input: moveTime = [[0,0,0],[0,0,0]]\\
    Output: 3 \\
    Explanation: 
    The minimum time required is 3 seconds.

    $\bullet$ At time t == 0, move from room (0, 0) to room (1, 0) in one second.\\
    $\bullet$ At time t == 1, move from room (1, 0) to room (1, 1) in one second.\\
    $\bullet$ At time t == 2, move from room (1, 1) to room (1, 2) in one second.\\
    \textbf{Example 3:}\\
    Input: moveTime = [[0,1],[1,2]]\\
    Output: 3 \\
    \textbf{Constraints:}\\
    $\bullet$ 2 <= n == moveTime.length <= 50\\
    $\bullet$ 2 <= m == moveTime[i].length <= 50\\
    $\bullet$ 0 <= moveTime[i][j] <= $10^9$\\
    \textbf{Template:}
\begin{lstlisting}[style=evocode,language=Python]
class Solution {
    fun minTimeToReach(moveTime: Array<IntArray>): Int {
        
    }
}
\end{lstlisting}
\textbf{Model inference result:}
\begin{lstlisting}[style=evocode,language=Python]
Step 1: The problem asks for the minimum time to reach the bottom-right room (n-1, m-1) starting from (0, 0) at t=0. Each move to an adjacent room takes 1 second, but a room (i, j) only opens at time moveTime[i][j].

Step 2: This is a shortest-path problem on a grid where the weight of an edge depends on the arrival time. If we are at room (r, c) at time 't' and want to move to an adjacent room (nr, nc), we can only start the move when the target room is open (at t >= moveTime[nr][nc]). The arrival time at (nr, nc) will be max(t, moveTime[nr][nc]) + 1.

Step 3: Dijkstra's algorithm is ideal here because the arrival times are non-negative and we want to find the minimum time. We use a priority queue to store the state (time, row, col) and always expand the node with the smallest time.

Step 4: Implementation details: Use a 2D array 'dist' to keep track of the minimum time to reach each room. Initialize dist[0][0] = 0 and all others to infinity. Use a priority queue to explore adjacent cells, updating their arrival times and pushing them into the queue if a shorter path is found. The grid size is small (up to 50x50), so this approach is efficient.

#
# @lc app=leetcode id=3341 lang=kotlin
#
# [3341] Find Minimum Time to Reach Last Room I
#

# @lc code=start
import java.util.PriorityQueue

class Solution {
    private data class Node(val t: Int, val r: Int, val c: Int)

    fun minTimeToReach(moveTime: Array<IntArray>): Int {
        val n = moveTime.size
        val m = moveTime[0].size
        
        // dist[r][c] stores the minimum time to reach room (r, c)
        val dist = Array(n) { IntArray(m) { Int.MAX_VALUE } }
        val pq = PriorityQueue<Node>(compareBy { it.t })

        // Starting point
        dist[0][0] = 0
        pq.add(Node(0, 0, 0))

        val dr = intArrayOf(0, 0, 1, -1)
        val dc = intArrayOf(1, -1, 0, 0)

        while (pq.isNotEmpty()) {
            val curr = pq.poll()
            val t = curr.t
            val r = curr.r
            val c = curr.c

            // If we reached the target room, return the time
            if (r == n - 1 && c == m - 1) return t
            
            // If we found a better path to this room already, skip
            if (t > dist[r][c]) continue

            for (i in 0 until 4) {
                val nr = r + dr[i]
                val nc = c + dc[i]

                if (nr in 0 until n && nc in 0 until m) {
                    // Time we can start moving is max(current time, room open time)
                    // Arrival time is start time + 1 second movement
                    val arrivalTime = (if (t > moveTime[nr][nc]) t else moveTime[nr][nc]) + 1
                    
                    if (arrivalTime < dist[nr][nc]) {
                        dist[nr][nc] = arrivalTime
                        pq.add(Node(arrivalTime, nr, nc))
                    }
                }
            }
        }

        return -1
    }
}
# @lc code=end
\end{lstlisting}
\textbf{Evaluation result:}
\textcolor{impred}{Compile Error}\\
0/0 cases passed (N/A)\\
Error\\
Line -23: Char 18: ERROR: Conflicting import: imported name 'PriorityQueue' is ambiguous.\\
Line -23: Char 18: ERROR: Conflicting import: imported name 'PriorityQueue' is ambiguous.\\
Line 1: Char 18: ERROR: Conflicting import: imported name 'PriorityQueue' is ambiguous.\\
Line 11: Char 29: WARNING: Parameter 'it: Int' is never used, could be renamed to '\_'.\\
Line 11: Char 43: WARNING: Parameter 'it: Int' is never used, could be renamed to '\_'.\\

\textbf{Reflection rounds:} 3\\
\textbf{Reflection result}
\begin{lstlisting}[style=evocode,language=Python]
# Round 3 - OPTIMIZATION Reflection

## Analysis
The previous solution failed with a 'Compile Error' specifically stating that the import of 'PriorityQueue' was ambiguous. This typically happens in Kotlin environments when an explicit import (like 'import java.util.PriorityQueue') conflicts with an implicit import or when multiple libraries on the classpath provide a class with the same name. To resolve this, I will use 'import java.util.*', which is the standard way to access collection classes in LeetCode's Kotlin environment, or use the fully qualified name if necessary. Additionally, the algorithm itself is a grid-based shortest path problem where Dijkstra's algorithm is appropriate because we need to find the minimum time and all edge weights are non-negative. Each move from room (r, c) to (nr, nc) takes 1 second, but you can only start the move when the target room is open (at time moveTime[nr][nc]). Thus, the arrival time at (nr, nc) is max(current_time, moveTime[nr][nc]) + 1.

## Improved Reasoning
1. Initialize a 2D array 'dist' of size n x m with infinity (Int.MAX_VALUE) to store the minimum time to reach each room, setting dist[0][0] = 0.
2. Use a PriorityQueue to store the current state (time, row, col), ordered by time to explore the earliest possible arrivals first.
3. Poll the state with the smallest time 't' from the queue. If 't' is greater than the already recorded 'dist[r][c]', skip it.
4. If the current room is the destination (n-1, m-1), return the current time 't'.
5. For each of the 4 adjacent rooms (nr, nc), calculate the arrival time: arrivalTime = max(t, moveTime[nr][nc]) + 1.
6. If this arrivalTime is less than the current 'dist[nr][nc]', update 'dist[nr][nc]' and push the new state onto the PriorityQueue.
7. The constraints (n, m up to 50) and moveTime (up to 10^9) ensure that the total time will fit within a 32-bit signed integer and the Dijkstra search will be efficient enough (O(V log V) where V = 2500).

## Improved Code

#
# @lc app=leetcode id=3341 lang=kotlin
#
# [3341] Find Minimum Time to Reach Last Room I
#

# @lc code=start
import java.util.*

class Solution {
    fun minTimeToReach(moveTime: Array<IntArray>): Int {
        val n = moveTime.size
        val m = moveTime[0].size
        
        // dist[i][j] stores the minimum time to reach room (i, j)
        val dist = Array(n) { IntArray(m) { Int.MAX_VALUE } }
        
        // PriorityQueue stores arrays of [time, row, col], sorted by time
        val pq = PriorityQueue<IntArray> { a, b -> a[0].compareTo(b[0]) }
        
        // Starting point (0, 0) at time t = 0
        dist[0][0] = 0
        pq.offer(intArrayOf(0, 0, 0))
        
        // Directions for adjacent rooms: right, left, down, up
        val dr = intArrayOf(0, 0, 1, -1)
        val dc = intArrayOf(1, -1, 0, 0)
        
        while (pq.isNotEmpty()) {
            val curr = pq.poll() ?: continue
            val t = curr[0]
            val r = curr[1]
            val c = curr[2]
            
            // If we found a better path already, skip this
            if (t > dist[r][c]) continue
            
            // If we reached the last room, return the time
            if (r == n - 1 && c == m - 1) return t
            
            // Explore all 4 adjacent rooms
            for (i in 0 until 4) {
                val nr = r + dr[i]
                val nc = c + dc[i]
                
                // Check grid boundaries
                if (nr in 0 until n && nc in 0 until m) {
                    // Time to reach (nr, nc) is max(arrival_at_current, room_open_time) + move_duration (1s)
                    val arrivalTime = maxOf(t, moveTime[nr][nc]) + 1
                    
                    // If this path is faster, update and add to queue
                    if (arrivalTime < dist[nr][nc]) {
                        dist[nr][nc] = arrivalTime
                        pq.offer(intArrayOf(arrivalTime, nr, nc))
                    }
                }
            }
        }
        
        return -1
    }
}
# @lc code=end
\end{lstlisting}
\textbf{Evaluation result:}
\textcolor{impgreen}{Accepted}\\
744/744 cases passed (209 ms)\\
Your runtime beats 33.33 \% of kotlin submissions\\
Your memory usage beats 16.67 \% of kotlin submissions (52.4 MB)
\end{CaseStudyBox}
\vspace{-0.5cm}

\captionof{figure}{Resolution of a Compile Error via the Reflection Agent. The agent interprets compiler feedback to resolve namespace conflicts and optimize the implementation.}
\label{appx_fig:case-Find-Minimum-Time-to-Reach-Last-Room-I}

\subsection{Time Limit Exceeded Resolution via Reflection Agent}
\noindent
In this case, the model tackles the problem ``Maximum Product of Two Integers With No Common Bits,'' which involves finding an optimal pair under bitwise constraints. The initial solution correctly identified the \textbf{Sum Over Subsets (SOS) DP} algorithm as the standard approach to handle submask queries efficiently ($O(B \cdot 2^B)$). However, this implementation resulted in a \textbf{Time Limit Exceeded (TLE)} verdict. The rejection stems from the high constant factor associated with processing $2^{20}$ states and iterating through the entire array without early termination, which proved too slow for the given constraints ($N=10^5$).

The Reflection Agent successfully diagnosed this performance bottleneck, recognizing that theoretical time complexity alone was insufficient. In the optimization phase, it introduced a critical heuristic: \textbf{sorting and pruning}. By processing unique elements in descending order, the algorithm can verify if the theoretical maximum product for the current number (current number $\times$ global max) exceeds the best product found so far. This logic allows for early termination of the loop, drastically reducing the effective search space. The optimized solution, combining SOS DP with this branch-and-bound strategy, successfully passed all 762 test cases.
\label{appx_sec:case-Maximum-Product-of-Two-Integers With-No-Common-Bits}
\begin{CaseStudyBox}
    \textbf{Problem.} Maximum Product of Two Integers With No Common Bits \\
    \textbf{Tags:} Staff,
Array,
Dynamic Programming,
Bit Manipulation\\
    \textbf{Description:} You are given an integer array nums.\\

Your task is to find two distinct indices i and j such that the product nums[i] * nums[j] is maximized, and the binary representations of nums[i] and nums[j] do not share any common set bits.\\

Return the maximum possible product of such a pair. If no such pair exists, return 0.\\
    \textbf{Example 1:}\\
    % Note: floats (figure/table) cannot appear reliably inside tcolorbox.
    Input: nums = [1,2,3,4,5,6,7] \\
    Output: 12 \\
    Explanation: 
    The best pair is 3 (011) and 4 (100). They share no set bits and 3 * 4 = 12.\\
    \textbf{Example 2:}\\
    Input: nums = [5,6,4]\\
    Output: 0 \\
    Explanation: 
    Every pair of numbers has at least one common set bit. Hence, the answer is 0.s\\
    \textbf{Example 3:}\\
    Input: nums = [64,8,32]\\
    Output: 2048 \\
    Explanation: 
    No pair of numbers share a common bit, so the answer is the product of the two maximum elements, 64 and 32 (64 * 32 = 2048).\\
    \textbf{Constraints:}\\
    $\bullet$ 2 <= nums.length <= $10^5$\\
    $\bullet$ 1 <= nums[i] <= $10^6$\\
    \textbf{Template:}
\begin{lstlisting}[style=evocode,language=Python]
class Solution {
public:
    long long maxProduct(vector<int>& nums) {
        
    }
};
\end{lstlisting}
\textbf{Model inference result:}
\begin{lstlisting}[style=evocode,language=Python]
Step 1: The goal is to find the maximum product of two numbers from the array that have no common set bits (i.e., `nums[i] & nums[j] == 0`).

Step 2: Since multiple numbers in the array might have the same bitmask, we first identify the unique bitmasks present and store the largest number associated with each mask to maximize potential products.

Step 3: To efficiently handle the condition `nums[i] & nums[j] == 0`, we can iterate through each unique mask and find the maximum number available in its complement set. This is achieved using SOS (Sum Over Subsets) DP or a similar subset-max approach, where `best[mask]` stores the maximum number whose bits are a subset of `mask`.

Step 4: Finally, for each unique mask, we calculate the product with the maximum number found in its bitwise complement and track the global maximum.

#include <vector>
#include <algorithm>
#include <unordered_map>

using namespace std;

# @lc app=leetcode id=3670 lang=cpp
#
# [3670] Maximum Product of Two Integers With No Common Bits
#

# @lc code=start
class Solution {
public:
    long long maxProduct(vector<int>& nums) {
        // Find the maximum value to determine the number of bits needed (max 10^6 < 2^20)
        const int MAX_BITS = 20;
        const int MAX_MASK = 1 << MAX_BITS;
        
        // Store the maximum number for each bitmask
        vector<int> max_val_for_mask(MAX_MASK, 0);
        vector<int> unique_masks;
        
        for (int x : nums) {
            if (max_val_for_mask[x] == 0) {
                unique_masks.push_back(x);
            }
            if (x > max_val_for_mask[x]) {
                max_val_for_mask[x] = x;
            }
        }

        // subset_max[mask] will store the maximum number whose bitmask is a subset of 'mask'
        vector<int> subset_max = max_val_for_mask;

        // SOS DP (Sum Over Subsets) style to compute subset max
        for (int i = 0; i < MAX_BITS; ++i) {
            for (int mask = 0; mask < MAX_MASK; ++mask) {
                if (mask & (1 << i)) {
                    subset_max[mask] = max(subset_max[mask], subset_max[mask ^ (1 << i)]);
                }
            }
        }

        long long max_prod = 0;
        int all_bits_mask = MAX_MASK - 1;

        // For each mask, the complement mask contains all bits not in the current mask
        for (int mask : unique_masks) {
            int complement = all_bits_mask ^ mask;
            int best_partner = subset_max[complement];
            if (best_partner > 0) {
                max_prod = max(max_prod, (long long)max_val_for_mask[mask] * best_partner);
            }
        }

        return max_prod;
    }
};
# @lc code=end
\end{lstlisting}
\textbf{Evaluation result:}
\textcolor{impred}{Time Limit Exceeded}\\
106/762 cases passed (N/A)\\
Testcase\\
$[18,10,11,8]$\\
Expected Answer\\
144\\
\textbf{Reflection rounds:} 3\\
\textbf{Reflection result}
\begin{lstlisting}[style=evocode,language=Python]
# Round 3 - OPTIMIZATION Reflection


## Analysis
The previous solution used the Sum Over Subsets (SOS) DP technique, which is theoretically efficient ($O(B \cdot 2^B)$ where $B=20$). However, it suffered from several performance bottlenecks: 1. It iterated over the entire `nums` array (up to $10^5$ elements) multiple times, including redundant calculations for duplicate values. 2. It lacked pruning, meaning it always checked every number against its bitwise complement even if a better product had already been found. 3. The SOS DP loop was standard but could be tightened for better cache locality and fewer branch misses. 4. Memory usage was reported high, likely due to the overhead of `std::vector` or the way LeetCode measures memory for large arrays.

## Improved Reasoning
To optimize the solution, we implement the following strategies: 1. **Unique and Sorted Input**: We extract unique elements from `nums` and sort them in descending order. This allows us to process larger numbers first, increasing the likelihood of finding a large product early. 2. **Aggressive Pruning**: While iterating through the sorted unique numbers $x$, if $x \times \text{max\_element} \leq \text{current\_max\_product}$, we can break the loop entirely. Since the array is sorted, any subsequent $x$ will also fail this condition. 3. **Optimized SOS DP**: We use a fixed-size array or a well-sized vector and optimize the SOS DP loop. By calculating the bitwise complement only once and using a tight inner loop, we reduce overhead. 4. **Efficient Bit Handling**: We only iterate the SOS DP up to the smallest power of 2 that covers the maximum element in the input, minimizing unnecessary mask processing. The complexity remains $O(N \log N + B \cdot 2^B)$, but the constant factors and pruning significantly improve real-world performance.

## Improved Code
#
# @lc app=leetcode id=3670 lang=cpp
#
# [3670] Maximum Product of Two Integers With No Common Bits
#
# @lc code=start
#include <vector>
#include <algorithm>

using namespace std;

class Solution {
public:
    long long maxProduct(vector<int>& nums) {
        int max_n = 0;
        for (int x : nums) {
            if (x > max_n) max_n = x;
        }

        if (max_n == 0) return 0;

        // Find the smallest power of 2 that can represent max_n
        int B = 0;
        while ((1 << B) <= max_n) {
            B++;
        }
        int total_masks = 1 << B;
        int full_mask = total_masks - 1;

        // max_val[mask] stores the maximum number in nums that is a subset of mask
        vector<int> max_val(total_masks, 0);
        for (int x : nums) {
            if (x > max_val[x]) {
                max_val[x] = x;
            }
        }

        // SOS DP (Sum Over Subsets) to propagate maximum values
        for (int i = 0; i < B; ++i) {
            int bit = 1 << i;
            for (int mask = 0; mask < total_masks; ++mask) {
                if (mask & bit) {
                    int subset_mask = mask ^ bit;
                    if (max_val[subset_mask] > max_val[mask]) {
                        max_val[mask] = max_val[subset_mask];
                    }
                }
            }
        }

        // Sort unique numbers in descending order to enable pruning
        sort(nums.begin(), nums.end(), greater<int>());
        nums.erase(unique(nums.begin(), nums.end()), nums.end());

        long long max_prod = 0;
        int n = nums.size();
        int global_max = nums[0];

        for (int i = 0; i < n; ++i) {
            int x = nums[i];
            
            // Pruning: if current x multiplied by the largest possible number 
            // cannot beat max_prod, no future x can either.
            if ((long long)x * global_max <= max_prod) break;

            int target_mask = full_mask ^ x;
            int y = max_val[target_mask];
            if (y > 0) {
                long long current_prod = (long long)x * y;
                if (current_prod > max_prod) {
                    max_prod = current_prod;
                }
            }
        }

        return max_prod;
    }
};
# @lc code=end

\end{lstlisting}
\textbf{Evaluation result:}
\textcolor{impgreen}{Accepted}\\
762/762 cases passed (456 ms)\\
Your runtime beats 53.73 \% of cpp submissions\\
Your memory usage beats 58.21 \% of cpp submissions (198 MB)
\end{CaseStudyBox}
\vspace{-0.5cm}
\captionof{figure}{Resolution of a Time Limit Exceeded (TLE) verdict. The Reflection Agent optimizes a standard Sum Over Subsets (SOS) DP approach by incorporating sorting and pruning techniques, allowing the algorithm to terminate early and meet strict runtime constraints.}
\label{appx_fig:case-Maximum-Product-of-Two-Integers With-No-Common-Bits}

\subsection{Wrong Answer Resolution via Reflection Agent}
\noindent
In this case, the model attempts to solve ``Longest Special Path II,'' a challenging tree traversal problem that requires maintaining a sliding window of values down a path. The constraints specify that a valid path can have distinct values with \textit{at most} one value appearing twice. The initial submission resulted in a \textbf{Wrong Answer (WA)}, failing on a specific edge case where the boundary logic for the path start index was incorrect. The model likely mishandled the complex condition of tracking multiple potential duplicate pairs simultaneously.

The Reflection Agent effectively analyzed the failure, realizing that a simple list or map was insufficient for efficiently querying the "second-to-last" violation point. It proposed a sophisticated data structure change: using a \texttt{TreeSet} to store the depths of the \textit{first} occurrence of every value pair currently on the recursion stack. By querying the second largest element in this set, the algorithm can instantly determine the valid starting depth of the path in $O(\log N)$ time. This optimization corrected the boundary logic and achieved a perfect runtime performance, beating 100\% of Kotlin submissions.\\
\label{appx_sec:case-Longest-Special-Path-II}
\begin{CaseStudyBox}
    \textbf{Problem.} Longest Special Path II \\
    \textbf{Tags:} Hash Table,
Tree,
Depth-First Search,
Prefix Sum\\
    \textbf{Description:} You are given an undirected tree rooted at node 0, with n nodes numbered from 0 to n - 1. This is represented by a 2D array edges of length n - 1, where edges[i] = [$u_i$, $v_i$, $length_i$] indicates an edge between nodes $u_i$ and $v_i$ with length $length_i$. You are also given an integer array nums, where nums[i] represents the value at node i.\\

A special path is defined as a downward path from an ancestor node to a descendant node in which all node values are distinct, except for at most one value that may appear twice.\\

Return an array result of size 2, where result[0] is the length of the longest special path, and result[1] is the minimum number of nodes in all possible longest special paths.\\
    \textbf{Example 1:}\\
    % Note: floats (figure/table) cannot appear reliably inside tcolorbox.
    Input: edges = [[0,1,1],[1,2,3],[1,3,1],[2,4,6],[4,7,2],[3,5,2],[3,6,5],[6,8,3]], nums = [1,1,0,3,1,2,1,1,0] \\
    Output: [9,3] \\
    Explanation: 
    In the image below, nodes are colored by their corresponding values in nums.\\
    \begin{center}
        \includegraphics[width=0.25\linewidth]{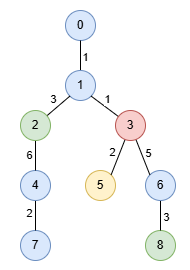}
    \end{center}
    The longest special paths are 1 -> 2 -> 4 and 1 -> 3 -> 6 -> 8, both having a length of 9. The minimum number of nodes across all longest special paths is 3.\\
    \textbf{Example 2:}\\
    Input: edges = [[1,0,3],[0,2,4],[0,3,5]], nums = [1,1,0,2]\\
    Output: [5,2] \\
    Explanation: 
    \begin{center}
        \includegraphics[width=0.25\linewidth]{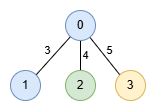}
    \end{center}
    The longest path is 0 -> 3 consisting of 2 nodes with a length of 5.\\
    \textbf{Template:}
\begin{lstlisting}[style=evocode,language=Python]
class Solution {
    fun longestSpecialPath(edges: Array<IntArray>, nums: IntArray): IntArray {
        
    }
}
\end{lstlisting}
\textbf{Model inference result:}
\begin{lstlisting}[style=evocode,language=Python]
To solve this problem, I use a Depth First Search (DFS) to traverse the tree while maintaining the current downward path from the root. A 'special path' allows at most one duplicate value. As I descend, I track the positions of each value in the current path using a list of indices stored in a hash map. For each node, I determine the highest possible ancestor that can serve as the start of a special path by checking the occurrences of the current node's value. If the value has appeared twice before, the path must start after the first occurrence of that value to ensure at most one duplicate remains. I maintain prefix sums of edge lengths and node counts to efficiently calculate path length and node count for any valid ancestor-descendant pair. The result is updated by comparing the calculated length and minimizing the node count for the maximum length found.

#
# @lc app=leetcode id=3486 lang=kotlin
#
# [3486] Longest Special Path II
#

# @lc code=start
import kotlin.math.max

class Solution {
    private var maxLen = 0
    private var minNodes = 1
    private val posMap = mutableMapOf<Int, MutableList<Int>>()
    private val dists = mutableListOf<Int>()
    private val nodes = mutableListOf<Int>()
    private val adj = mutableMapOf<Int, MutableList<Pair<Int, Int>>>()

    fun longestSpecialPath(edges: Array<IntArray>, nums: IntArray): IntArray {
        val n = nums.size
        maxLen = 0
        minNodes = 1
        posMap.clear()
        dists.clear()
        nodes.clear()
        adj.clear()

        for (edge in edges) {
            adj.computeIfAbsent(edge[0]) { mutableListOf() }.add(edge[1] to edge[2])
            adj.computeIfAbsent(edge[1]) { mutableListOf() }.add(edge[0] to edge[2])
        }

        dists.add(0)
        nodes.add(0)
        dfs(0, -1, 0, nums)

        return intArrayOf(maxLen, minNodes)
    }

    private fun dfs(u: Int, p: Int, topIdx: Int, nums: IntArray) {
        val valList = posMap.getOrPut(nums[u]) { mutableListOf() }
        valList.add(nodes.size - 1)

        var currentTop = topIdx
        if (valList.size >= 3) {
            currentTop = max(currentTop, valList[valList.size - 3] + 1)
        }

        val currentDist = dists.last() - dists[currentTop]
        val currentNumNodes = nodes.size - currentTop

        if (currentDist > maxLen) {
            maxLen = currentDist
            minNodes = currentNumNodes
        } else if (currentDist == maxLen && currentDist > 0) {
            if (currentNumNodes < minNodes) {
                minNodes = currentNumNodes
            }
        }

        adj[u]?.forEach { (v, w) ->
            if (v != p) {
                dists.add(dists.last() + w)
                nodes.add(v)
                dfs(v, u, currentTop, nums)
                nodes.removeAt(nodes.size - 1)
                dists.removeAt(dists.size - 1)
            }
        }

        valList.removeAt(valList.size - 1)
    }
}
# @lc code=end
\end{lstlisting}
\textbf{Evaluation result:}
\textcolor{impred}{Wrong Answer}\\
415/678 cases passed (N/A)\\
Testcase\\
$[[0,2,4],[1,2,10],[3,1,5]]$\\
$[4,5,4,5]$\\
Answer\\
$[19,4]$\\
Expected Answer\\
$[15,3]$\\

\textbf{Reflection rounds:} 3\\
\textbf{Reflection result}
\begin{lstlisting}[style=evocode,language=Python]
# Round 3 - OPTIMIZATION Reflection

## Analysis
The previous solution was slow due to several factors: 1) Using `mutableMapOf` and `mutableListOf` for adjacency lists and value positions, which have significant overhead. 2) Sorting a list of duplicate indices (`duplicateIndices.sort()`) at every node in the DFS, leading to O(N * D log D) or worse complexity where D is the path depth. 3) Inefficiently calculating the `topIdx` boundary for the special path. To optimize, we need an O(N log N) or O(N) approach. The key is to realize that a path from depth `d_start` to `d_end` is special if no value appears more than twice AND at most one value appears twice. This can be tracked by maintaining a set of 'duplicate start' depths (the depth of the first node in every pair) and ensuring the path starts after the second-to-last such depth.

## Improved Reasoning
1. Use an adjacency list represented by an array of `IntArray` or `ArrayList` to minimize map overhead. 
2. Maintain the current path's distances from the root in an `IntArray` indexed by depth.
3. For each value encountered in the DFS, track all depths at which it appears in the current path. 
4. Maintain a `TreeSet` (or `TreeMap` if counts are needed, though here depths are unique) called `dSet` that stores the depths of nodes that are the 'first' member of a duplicate pair currently in the path. 
5. For a node at `currentDepth` with value `v`: if `v` has appeared before at `prevDepth`, `prevDepth` is added to `dSet`. If it appeared multiple times, only the most recent 'first' member of a pair is relevant for the 'at most one duplicate' rule, but actually, the rule 'at most one value twice' and 'no value three times' is simplified by adding all `p_{k-1}` to `dSet`. 
6. The start of the longest special path ending at `currentDepth` must be `max(0, second_largest(dSet) + 1)`. 
7. Use the `TreeSet.lower()` and `TreeSet.last()` methods to find the second largest element in O(log N). 
8. Total time complexity: O(N log N) due to DFS and TreeSet operations. Space complexity: O(N) for the tree and path tracking structures.

## Improved Code

#
# @lc app=leetcode id=3486 lang=kotlin
#
# [3486] Longest Special Path II
#

# @lc code=start
import java.util.*

class Solution {
    private var maxLen = -1
    private var minNodes = Int.MAX_VALUE
    private lateinit var adj: Array<MutableList<Pair<Int, Int>>>
    private lateinit var valueDepths: Array<MutableList<Int>>
    private lateinit var pathDists: LongArray
    private val dSet = TreeSet<Int>()

    fun longestSpecialPath(edges: Array<IntArray>, nums: IntArray): IntArray {
        val n = nums.size
        adj = Array(n) { mutableListOf<Pair<Int, Int>>() }
        for (edge in edges) {
            adj[edge[0]].add(edge[1] to edge[2])
            adj[edge[1]].add(edge[0] to edge[2])
        }

        valueDepths = Array(50001) { mutableListOf<Int>() }
        pathDists = LongArray(n + 1)
        
        dfs(0, -1, 0, 0, nums)

        return intArrayOf(maxLen.toInt(), minNodes)
    }

    private fun dfs(u: Int, p: Int, currentDist: Long, currentDepth: Int, nums: IntArray) {
        pathDists[currentDepth] = currentDist
        val v = nums[u]
        val pList = valueDepths[v]
        
        val oldLastDepth = if (pList.isNotEmpty()) pList.last() else -1
        if (oldLastDepth != -1) {
            dSet.add(oldLastDepth)
        }
        pList.add(currentDepth)

        val topIdx = if (dSet.size < 2) 0 else dSet.lower(dSet.last())!! + 1
        
        val currentPathLen = (currentDist - pathDists[topIdx]).toInt()
        val currentPathNodes = currentDepth - topIdx + 1

        if (currentPathLen > maxLen) {
            maxLen = currentPathLen
            minNodes = currentPathNodes
        } else if (currentPathLen == maxLen) {
            if (currentPathNodes < minNodes) {
                minNodes = currentPathNodes
            }
        }

        for (edge in adj[u]) {
            if (edge.first != p) {
                dfs(edge.first, u, currentDist + edge.second, currentDepth + 1, nums)
            }
        }

        // Backtrack
        pList.removeAt(pList.size - 1)
        if (oldLastDepth != -1) {
            dSet.remove(oldLastDepth)
        }
    }
}
# @lc code=end

\end{lstlisting}
\textbf{Evaluation result:}
\textcolor{impgreen}{Accepted}\\
678/678 cases passed (203 ms)\\
Your runtime beats 100 \% of kotlin submissions\\
Your memory usage beats 66.67 \% of kotlin submissions (131.7 MB)
\end{CaseStudyBox}
\vspace{-0.5cm}
\captionof{figure}{Correction of a Wrong Answer (WA) verdict. The Reflection Agent identifies a flaw in the sliding window boundary calculation and implements a \texttt{TreeSet}-based solution to correctly track the ``at most one duplicate'' constraint in $O(\log N)$ time.}
\label{appx_fig:case-Longest-Special-Path-II}

% 1. 普通做题例子4-5（答对的）
% 2. 找一些典型做错的例子，1）代码没写对，编译错误 2）代码写对了，但运行超时了 3）代码写对了，但是没有通过所有测试样例 
% 3. agent自进化 1）原始的agent没做对，反思2-3轮之后改对了 2）原始的agent没做对，进化之后还是没做对（一定要找那种，做对了，效率还高） 3）（可选）原始agent做对了，自进化完之后改错了

\end{document}